\documentclass[english,11pt,aps,prd,a4paper,preprintnumbers,nofootinbib,showpacs,superscriptaddress, notitlepage]{revtex4-1} 
\pdfoutput=1
\usepackage[utf8]{inputenc}
\usepackage{algorithm}
\usepackage{mathtools,amssymb}
\usepackage{calc}
\usepackage{appendix}
\usepackage{natbib}
\usepackage{bm}
\usepackage{graphics}
\usepackage{graphicx}
\usepackage[usenames,dvipsnames]{color}  
\usepackage{caption}
\usepackage{stackengine}
\usepackage{subcaption}
\usepackage[export]{adjustbox} 
\usepackage{amsmath,lipsum}
\usepackage{amssymb}
\usepackage[colorlinks=true,citecolor=darkred,urlcolor=darkred, pdfborder={0 0 0}]{hyperref}
\usepackage[normalem]{ulem}
\usepackage{braket}
\usepackage{IEEEtrantools}
\usepackage{tikz-feynman}
\tikzfeynmanset{compat=1.0.0} 

\usepackage{comment}
\usepackage{placeins}

\definecolor{darkred}{rgb}{0.6,0,0}

\definecolor{linkcolor}{rgb}{0,0,0.5}

\usepackage{braket}
\definecolor{linkcolor}{rgb}{0,0,0.5}
\allowdisplaybreaks
\allowdisplaybreaks[1]
\allowdisplaybreaks[2]

\def\gsim{\raise0.3ex\hbox{$\;>$\kern-0.75em\raise-1.1ex\hbox{$\sim\;$}}}
\def\lsim{\raise0.3ex\hbox{$\;<$\kern-0.75em\raise-1.1ex\hbox{$\sim\;$}}}
\def\beqn#1{\begin{equation}\label{#1}}
\def\eeqn{\end{equation}}

\def\beqa#1{\begin{eqnarray}\label{#1}}
\def\eeqa{\end{eqnarray}}

\def\0nbb {$0\nu\beta\beta$ }

\def\Z2{$\mathcal{Z}_2$}
\def\A4{${A_4}$}

\def\red{\color{red}{}}

\newcommand {\ignore}[1]{}

\def\321{$\mathrm{SU(3) \otimes SU(2) \otimes U(1)}$ }

\newcommand{\AddrIISERB}{Department of Physics, Indian Institute of Science Education and Research - Bhopal, \\ 
Bhopal Bypass Road, Bhauri, Bhopal 462066, India}

\begin{document}

 \bibliographystyle{unsrt}   

\title{\color{BrickRed}
Cutting the Scotogenic loop: Adding flavor to Dark Matter \\
}

\author{Ranjeet Kumar}\email{ranjeet20@iiserb.ac.in}
\affiliation{\AddrIISERB}
\author{Newton Nath} 
\email{newton.nath@ific.uv.es}
\affiliation{Instituto de Física Corpuscular, CSIC-Universitat de València, and Departament de Física Teòrica, Universitat de València, C/Catedrático José Beltrán 2, Paterna 46980, Spain}
\affiliation{Istituto Nazionale di Fisica Nucleare,   Via  Orabona  4,  70126  Bari, Italy} 
\author{Rahul Srivastava}\email{rahul@iiserb.ac.in}
\affiliation{\AddrIISERB}

\begin{abstract}
\vspace{1cm}
\noindent

We introduce a framework for hybrid neutrino mass generation, wherein scotogenic dark sector particles, including dark matter, are charged non-trivially under the  $A_4$ flavor symmetry. The spontaneous breaking of the \A4 group to residual \Z2 subgroup results in the ``cutting" of the radiative loop.  As a consequence the neutrinos acquire mass through the hybrid ``scoto-seesaw" mass mechanism, combining aspects of both the tree-level seesaw and one-loop scotogenic mechanisms, with the residual \Z2 subgroup ensuring the stability of the dark matter.
The flavor symmetry also leads to several predictions including the normal ordering of neutrino masses and ``generalized $\mu-\tau$ reflection symmetry" in leptonic mixing.  Additionally, it gives testable predictions for neutrinoless double beta decay and a lower limit on the lightest neutrino mass.
Finally, $A_4 \to$ \Z2 breaking also leaves its imprint on the dark sector and ties it with the neutrino masses and mixing.
The model allows only scalar dark matter, whose mass has a theoretical upper limit of $\lsim$ 600 GeV, with viable parameter space satisfying all dark matter constraints, available only up to about 80 GeV.
Conversely, fermionic dark matter is excluded due to constraints from the neutrino sector. Various aspects of this highly predictive framework can be tested in both current and upcoming neutrino and dark matter experiments.
\end{abstract}

\maketitle

\section{Introduction}
\label{sec:introduction}

Despite its remarkable success in elucidating various observed natural phenomena, the Standard Model (SM)  grapples with several unresolved questions. A notable limitation lies in its inability to explain the experimentally observed non-zero neutrino masses at a renormalizable level and their mixing patterns, as observed in solar and atmospheric neutrino oscillation experiments~\cite{SNO:2001kpb,Super-Kamiokande:1998kpq}.
There are several proposals in the literature to explain the tiny masses of neutrinos. Although these models provide an understanding of the smallness of neutrino masses through various mass mechanisms, they do not shed light on the mixing and flavor structure of the leptonic sector. New flavor symmetries are often used to understand the leptonic (and/or quark) mixing structure. This flavor symmetry approach has proven highly successful in elucidating and predicting the flavor structure. Notably, the discret non-abelian $A_4$ symmetry~\cite{Ma:2001dn, Babu:2002dz, Altarelli:2005yx} has been the most popular flavor group used for such purpose. However, the flavor symmetric approach often fails to incorporate the existence of the two different mass scales namely, the atmospheric mass-squared difference, $\Delta m^2_{\rm {atm}}$,  and the solar mass-squared difference, $\Delta m^2_{\rm {sol}}$, observed in neutrino oscillation experiments~\cite{Capozzi:2021fjo, deSalas:2020pgw, Esteban:2020cvm}. 
%

Apart from the neutrino sector, the absence of a viable candidate for cosmological dark matter (DM) in the SM raises another significant concern. DM is typically expected to be electrically neutral, non-baryonic particle(s), and the recent Planck data indicates that it constitutes approximately 85\% of the observed matter in the Universe~\cite{Planck:2018vyg}. Thus, any beyond the SM (BSM) extension should address both these shortcomings.
\begin{figure}[h!]
    \centering
      \includegraphics[width=0.50\textwidth]{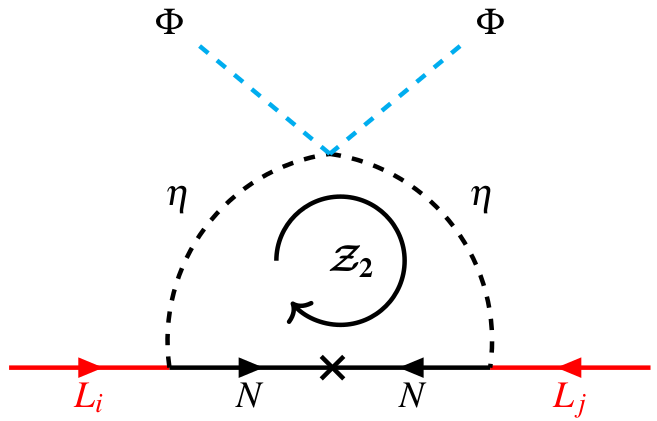}
    \caption{\footnotesize  Neutrino mass generation at the one-loop level in the canonical scotogenic model. Note that here a new dark \Z2 symmetry is needed to ensure DM stability. }
    \label{fig:UVloop}
    \end{figure}
The ``scotogenic'' mechanism, originally proposed in \cite{Ma:2006km}, remains one of the simplest and most elegant approaches to establish a possible connection between the DM and neutrino mass generation. The hallmark of the scotogenic framework is the loop-level generation of neutrino masses through ``dark sector" particles running in the loop as shown in Fig.~\ref{fig:UVloop}, with the lightest dark sector particle being the DM. 
The scotogenic mechanism and its variants have received much attention in recent times~\cite{Ma:2008cu,Hirsch:2013ola,Merle:2016scw,Diaz:2016udz,Choubey:2017yyn,Bonilla:2018ynb,CentellesChulia:2019gic,Srivastava:2019xhh,CentellesChulia:2019xky,Restrepo:2019ilz,Avila:2019hhv,
Kang:2019sab,Leite:2019grf,Borah:2019aeq,CarcamoHernandez:2020ehn,Leite:2020bnb,Leite:2020wjl}.
However, it is worth pointing out here that the canonical scotogenic model~\cite{Ma:2006km}  requires an \textit{ad-hoc} $\mathcal{Z}_2$ symmetry, as shown in Fig.~\ref{fig:UVloop}, to ensure the stability of DM. The various variants of the scotogenic model known in the literature, also require a new symmetry to ensure the stability of the DM. Furthermore, like other neutrino mass mechanisms, it also fails to explain the flavor structure of the lepton sector or the presence of two different mass scales in neutrino oscillations.
    
Recently, in~\cite{Rojas:2018wym}, an attempt has been made to successfully explain the origin of the two mass-squared differences observed in neutrino oscillation data while preserving the main features of the scotogenic model.
This is achieved through a  ``hybrid mass mechanism'' which is usually referred to as the ``scoto-seesaw'' mechanism\footnote{Note that the mass hierarchy among neutrinos is milder compared to that among different generations of other SM fermions. However, since $\frac{\Delta m^2_{\rm{sol}}}{\Delta m^2_{\rm{atm}}} \, \approx 10^{-2}$, this difference can have a natural explanation in terms of a hybrid mass generation mechanism such as ``scoto-seesaw".} where the atmospheric scale arises from the tree-level seesaw, whereas the solar scale has a radiative scotogenic origin~\cite{Rojas:2018wym,Barreiros:2020gxu,Mandal:2021yph,Barreiros:2022aqu, Ganguly:2023jml,Ganguly:2022qxj,VanDong:2023xbd,Leite:2023gzl,Kumar:2023moh}. However, scoto-seesaw like other neutrino mass mechanisms, still requires the addition of a flavor symmetry to explain the leptonic mixing pattern~\cite{Barreiros:2020gxu,Barreiros:2022aqu,Kumar:2023moh}. In summary, usually, even the most simple and elegant BSM models require the addition of at least a new dark as well as a flavor symmetry, along with the associated expansion of the BSM particle content, to account for DM and to explain neutrino mass and flavor structure.

In this work, we develop a simple framework with a very minimal particle content, that can explain the DM stability, neutrino mass generation, and flavor structure of the lepton sector along with the two mass scales of neutrino oscillation experiments using only a single flavor symmetry. Within this framework, breaking of flavor symmetry induces a ``scoto-seesaw'' \cite{Rojas:2018wym} like scenario, and DM stability is ensured by one of its unbroken subgroups.
We demonstrate this idea by explicitly constructing a simple model with minimal particle content. 
We start by imposing the $A_4$ flavor symmetry on a scotogenic-like radiative loop as shown schematically in the left panel of Fig.~\ref{fig:breakingA4}. 
It is important to highlight that, unlike the $\mathcal{Z}_2$ symmetry of the scotogenic mechanism~\cite{Ma:2006km}, which serves solely to stabilize DM candidate, the $A_4$ symmetry not only generates the flavor structure but also includes subgroups that fulfill the stabilizing role of $\mathcal{Z}_2$. This makes $A_4$ a more comprehensive symmetry for achieving broader theoretical purposes. Such a flavor group can be connected to more fundamental groups like $SO(3)$~\cite{Grimus:2011fk}. Additionally, it has been recently proposed that such a group could originate from finite modular groups $\Gamma_N$ inspired by string theory and extra-dimensional models~\cite{deAdelhartToorop:2011re}.

As shown in Fig. \ref{fig:breakingA4},  both particles i.e. $\eta$ and $N$ running in the loop (left panel) transform as triplets under $A_4$ symmetry. 
The $A_4$ symmetry is then broken spontaneously by the vacuum expectation value (VEV) of the scalar $\eta$ field in such a way as to leave a \Z2 subgroup unbroken. The spontaneous breaking of $A_4$ symmetry to its residual $\mathcal{Z}_2$ subgroup amounts to cutting the radiative loop and leads to a hybrid scoto-seesaw mechanism as shown in the right panel of Fig.~\ref{fig:breakingA4}. 
The resulting scoto-seesaw mechanism implies that the neutrino masses are generated through contributions from both the tree-level type-I seesaw and the one-loop level scotogenic mechanism thus providing a natural explanation for the two different mass-squared differences observed in neutrino oscillation experiments.
The unbroken \Z2 symmetry plays the role of the scotogenic \textit{dark} symmetry. The lightest $\mathcal{Z}_2$ odd particle is then automatically stable and can be a viable DM candidate. 

 In the leptonic sector,  the $A_4$ flavor symmetry predicts a small range with a lower limit on the lightest neutrino mass, normal ordering of neutrino masses, and a ``generalized $\mu-\tau$ reflection symmetry" for the neutrino mixing parameters that are in good agreement with recent data\footnote{It's important to note that the ``$\mu-\tau$ reflection symmetry'', initially introduced in \cite{Harrison:2002et}, anticipates maximal values for both the Dirac CP phases and the atmospheric mixing angle. However, given that the best-fit values from neutrino oscillation data tend to favor non-maximal values for these parameters, the framework of ``generalized $\mu-\tau$ reflection symmetry" \cite{Chen:2015siy} emerges as a successful model, showcasing its predictive capability.}.
We find testable predictions for the neutrino oscillations, neutrinoless double beta $(0\nu\beta\beta)$ decay experiments along with beta decay experiments.
Finally, the imprints of the emergence of the dark symmetry from the flavor symmetry can be found in the dark sector as well.
For example, the particles within the dark sector in our model are constrained by an upper limit on their masses $\lsim$ 600 GeV. The fermionic DM case is completely ruled out by the constraints coming from the neutrino oscillation data. The scalar DM is also heavily constrained with the viable parameter space consistent with both neutrino and DM experimental results available only up to 80 GeV DM mass.

We structure the article as follows. In Sec.~\ref{sec:Model}, we describe the general framework of our model. In Sec.~\ref{sec:scalarsec} and \ref{sec:scalarsec}, we present the potential and mass spectrum of the scalar sector, respectively. Sec.~\ref{sec:NuSector} discusses the generation of neutrino mass along with the neutrino flavor structure. In Sec.~\ref{sec:Nuprediction}, we explain our numerical results for the neutrino sector.
Our quantitative results for the dark sector are presented in Sec.~\ref{sec:DarkSector}. 
Finally, we give concluding remarks in Sec.~\ref{sec:Conclusion}. The $A_4$ algebra and expanded form of scalar potential are given in Appendix~\ref{app:A4}, while detailed information regarding the fermionic DM case is provided in Appendix~\ref{app:fermionicDM}. In Appendix~\ref{app:OtherRe} we discuss different choices of $A_4$ transformation of the leptonic doublets. The Feynman diagrams relevant to relic abundance and direct detection computations are in Appendix~\ref{app:DM-DD}.

\section{Model set-up}\label{sec:Model}
We utilize a scotogenic-like radiative seesaw model in conjunction with \A4 flavor symmetry. 
In addition to being a popular flavor symmetry group, \A4 also has a \Z2 subgroup which can potentially serve as the dark symmetry as we discuss later~\cite{Hirsch:2010ru,Boucenna:2011tj,DeLaVega:2018bkp,Bonilla:2023pna}.
The field content and their transformations under the SM gauge group $SU(3)_C \otimes SU(2)_L \otimes U(1)_Y$ in addition to \A4 symmetry are provided in Table \ref{tab:FieldCharge}. Notice that in Appendix~\ref{app:OtherRe}, we discuss various choices of $A_4$ transformations for the leptonic doublets. Contrary to the findings in \cite{Bonilla:2023pna}, we highlight that the choice of $A_4$ transformations play a crucial role in predicting neutrino phenomenology.
\begin{center}
\begin{table}[h] 
\begin{tabular}{ |c||c|c| } 
 \hline
 Fields & $SU(3)_C\otimes SU(2)_L\otimes U(1)_Y$ & {\red \A4}  ${\color{Mahogany} \mathbf{\to}}$ \Z2 \\
 \hline \hline
$L_{i}$ & (1, 2, -1) &  {\red(1, $1'$, $1''$)} ${\color{Mahogany} \mathbf{\to}}$ ($\bm{+}$, $\bm{+}$, $\bm{+}$)  \\
$e_{R_i}$ & (1, 1, -2) &  {\red(1, $1'$, $1''$)} ${\color{Mahogany} \mathbf{\to}}$ ($\bm{+}$, $\bm{+}$, $\bm{+}$) \\
$\Phi$ & (1, 2, 1) &  {\red 1} ${\color{Mahogany} \mathbf{\to}}$ $\bm{+}$\\
\hline \hline
$\eta$ & (1, 2, 1) &  {\red 3} ${\color{Mahogany} \mathbf{\to}}$ ($\bm{+}$, $\bm{-}$, $\bm{-}$) \\
 N & (1, 1, 0) &  {\red 3} ${\color{Mahogany} \mathbf{\to}}$ ($\bm{+}$, $\bm{-}$, $\bm{-}$) \\
 \hline
\end{tabular}
\caption{\footnotesize Field content and their transformation properties under the  SM gauge group $ SU(3)_C\otimes SU(2)_L\otimes U(1)_Y$ and $A_4$ symmetry, where $i = 1, 2, 3$ represents generation indices. The \A4 group is spontaneously broken down to its \Z2 subgroup by the VEV of the $\eta$ scalar. Transformation properties of the residual \Z2 symmetry are listed in the last column.}
\label{tab:FieldCharge}
\end{table}
\end{center}
\vspace{-1cm}

Apart from SM particles, we introduce two types of BSM particles:  $SU(2)_L$ doublet scalars $\eta$ and singlet fermions $N$, both transforming as triplets under \A4 symmetry, see Appendix~\ref{sec:app-A4} for more details of \A4 flavor symmetry.
All SM particles transform as singlets under \A4 symmetry: the Higgs doublet $\Phi$ and all quarks transform as trivial singlet (1), and the charged leptons $L_i$ and $e_{R_i}$; $i = 1,2,3$ transform as singlets $(1, 1', 1'')$ under the \A4 symmetry.
The charge assignment in Table~\ref{tab:FieldCharge} ensures that, after the \A4 symmetry breaking, all SM particles are even (+) under \Z2, given their singlet nature under \A4 symmetry. The components of \A4 triplets $\eta,\ N$ split into two, the first components transforming as even (+) while the other components have odd (-) charges under \Z2, see Appendix~\ref{sec:app-A4} for more details. Thus the particle content of our model is divided into even and odd particles under residual \Z2 symmetry as shown in the last column of Table~\ref{tab:FieldCharge}. The odd particles will eventually become dark sector particles as discussed in Sec.~\ref{sec:DarkSector}.

\subsection{Scalar sector} \label{sec:scalarsec}

We start with the scalar sector of our model and discuss how the spontaneous breaking of \A4 symmetry leads to the emergence of residual \Z2 dark symmetry. The $SU(3)_C\otimes SU(2)_L\otimes U(1)_Y \otimes A_4$ invariant scalar potential is given as follows:
%
\begin{align}
V&=\mu_{\Phi}^2 \Phi^\dagger \Phi 
+ \mu_\eta^2 \left[\eta^\dagger\eta \right]_1 + \lambda_1 \left(\Phi^\dagger \Phi \right)^2+\lambda_2 \left[\eta^\dagger\eta \right]_1^2 +\lambda_3 \left[\eta^\dagger\eta \right]_{1^{\prime}}\left[\eta^\dagger\eta \right]_{1^{\prime\prime}} +\lambda_4 \left[\eta^\dagger\eta^\dagger \right]_{1^\prime}\left[\eta\eta \right]_{1^{\prime\prime}} \nonumber \\
&+\lambda_{4'}\left[\eta^\dagger\eta^\dagger \right]_{1^{\prime\prime}}\left[\eta\eta \right]_{1^\prime} 
+\lambda_5\left[\eta^\dagger\eta^\dagger \right]_{1}\left[\eta\eta \right]_{1} 
+\lambda_6\left(\left[\eta^\dagger \eta \right]_{3_{1}}\left[\eta^\dagger \eta \right]_{3_{1}}+h.c.\right)
+ \lambda_7 \left[\eta^\dagger \eta \right]_{3_{1}}\left[\eta^\dagger \eta \right]_{3_{2}} \nonumber \\
&+\lambda_8 \left[\eta^\dagger \eta^\dagger \right]_{3_{1}}\left[\eta \eta \right]_{3_{2}} 
+\lambda_9 \left[\eta^\dagger \eta \right]_1 \left(\Phi^\dagger \Phi \right) 
+\lambda_{10}\left[\eta^\dagger \Phi \right]_3 \left[\Phi^\dagger \eta \right]_3 +\lambda_{11}\left(\left[\eta^\dagger\eta^\dagger \right]_{1}\left(\Phi \Phi \right)+h.c.\right) \nonumber \\
&+\lambda_{12}\left(\left[\eta^\dagger\eta^\dagger \right]_{3_{1}}\left[\eta \Phi \right]_3+h.c.\right) 
+\lambda_{13}\left(\left[\eta^\dagger\eta^\dagger \right]_{3_{2}}\left[\eta \Phi \right]_3+h.c.\right) 
+\lambda_{14}\left(\left[\eta^\dagger \eta \right]_{3_{1}}\left[\eta^\dagger \Phi \right]_3+h.c.\right) \nonumber \\
&+\lambda_{15}\left(\left[\eta^\dagger \eta \right]_{3_{2}} \left[\eta^\dagger \Phi \right]_3+h.c.\right) \;.
\label{eq:scalarpot}
\end{align}
%
Where $[...]_p$; $p=1,1',1'',3,3_1,3_2$ denote the \A4 transformation of enclosed fields. Furthermore, $3_1, \ 3_2$ denote the two possible \A4 triplet contractions, see Eqs.~\eqref{eq:a4mrule} and~\eqref{eq:pr} of Appendix~\ref{sec:app-A4}.

%
Here it is important to emphasize a noteworthy aspect resulting from the \A4 symmetry.  Specifically, terms of the form $\eta^3 \Phi$ are permitted by the \A4 symmetry. However in canonical scotogenic model~\cite{Ma:2006km} as well in its variants where the \Z2 symmetry is explicitly added, they are forbidden due to the \Z2 odd (even) charge assignment of $\eta ~(\Phi)$ fields. The presence of such terms in our model would eventually lift the mass degeneracy between the components of the \A4 triplet $\eta$ after the symmetry breaking. The expanded form of this potential in terms of the component fields is given in Appendix~\ref{sec:app-scalar}.

Now coming to the breaking of \A4 symmetry, we aim to spontaneously break the \A4 symmetry in such a way that the \Z2 subgroup remains unbroken. This can be accomplished by giving VEV to the \A4 triplet scalar $\eta$ with the VEV alignment $\langle \eta \rangle = \frac{1}{\sqrt{2}}(v_2, 0, 0)$. Furthermore, after $A_4 \to \mathcal{Z}_2$ breaking, the components $\eta_i$; $i = 1,2,3$ split into two types under the residual \Z2 symmetry, transforming as
\begin{equation}
\eta_1 \to +\eta_1 \, , \quad \eta_{2} \to -\eta_{2}, \quad \eta_{3} \to -\eta_{3}\;
\end{equation}
while $\Phi$ being a \A4 singlet transform as $\Phi \to +\Phi$ under the residual \Z2.  The transformations of various representations of \A4 under the \Z2 subgroup are given in Appendix~\ref{sec:app-A4}. 

Since $\eta$ is a $SU(2)_L$ doublet, its VEV also breaks the electroweak symmetry along with the VEV of $ \Phi $.  Thus, the scalars have the following VEV alignments:
\begin{equation} \label{eq:vev}
\langle \Phi \rangle = \frac{v_1}{\sqrt{2}} ,\,  \langle \eta_1 \rangle =  \frac{v_2}{\sqrt{2}} \ {\rm and} \ \langle \eta_2 \rangle = \langle \eta_3 \rangle \, = 0  \,. 
\end{equation}
We define $ \tan \beta = v_2/v_1 $ and we will take $\tan \beta \sim 1$ in the rest of the analysis.
The potential is also CP conserving if all couplings are taken to be real and  $\lambda_{4^\prime}=\lambda_4$.
The requirement of perturbativity  leads to the constraints on the $\lambda_i$ couplings and $\tan\beta$ given by:
\begin{align} \label{eq:perturbcn} 
  \lambda_i \lsim \sqrt{4 \pi}, \quad i = 1,...,15, \quad  \tan\beta > 0.5 \, .
\end{align}
The tree-level stability of the vacuum can be ensured by the following conditions~\cite{Boucenna:2011tj}:
\begin{align} 
&\lambda_1>0 \hspace{0.05cm},  \hspace{0.5cm} \lambda_2+\lambda_3+2\lambda_4+\lambda_5>0 \hspace{0.05cm},\nonumber \\
&\lambda_1+3(\lambda_2+\lambda_3+2\lambda_4+\lambda_5)+3(\lambda_9+Q_1)+3(2\lambda_2-\lambda_3+\lambda_8+Q_2)-6\zeta >0,
\label{eq:vacuumstab}
\end{align}
where $Q_1 = {\rm Min }(\lambda_{10}-2|\lambda_{11}|,\ 0),\ Q_2 = {\rm Min }(\lambda_7-2|\lambda_4-\lambda_5-\lambda_6|, \ 0)$ and $\zeta=|\lambda_{12}|+|\lambda_{13}|+|\lambda_{14}|+|\lambda_{15}|.$ 
The last line of Eq.~\eqref{eq:vacuumstab} is an approximate condition for the stability which can be obtained following the procedure discussed in~\cite{Grzadkowski:2009bt,Buskin:2021eig,Emmanuel-Costa:2016vej}.

The $SU(2)_L$ doublet scalars $\Phi$ and $\eta_i$ after SSB, are expressed as: 
\begin{align} \label{eq:doubletexp}
\Phi & =
\begin{pmatrix}
\phi^{+}\\
(v_1+\phi^{ 0}+i\sigma_1^{ 0})/\sqrt{2}
\end{pmatrix},\, \quad 
\eta_1 = 
\begin{pmatrix}
\eta_1^{ +}\\
(v_2+\eta^{ R }_{1}+i\eta^{I}_{1})/\sqrt{2}
\end{pmatrix} ,\,
 \nonumber \\
\eta_2 & =
\begin{pmatrix}
\eta^{ +}_2\\
(\eta^{ R }_{2}+i\eta^{I}_{2})/\sqrt{2}
\end{pmatrix} ,\, \hspace{1.2cm}
\eta_3 =
\begin{pmatrix}
\eta^{+}_3\\
(\eta^{ R}_{3}+i\eta^{ I}_{3})/\sqrt{2}
\end{pmatrix}.
\end{align}
Note that $\eta_2 $ and $\eta_3$ transform as (-) under the residual \Z2 subgroup and have not acquired any VEV, ensuring \Z2 survives as a residual symmetry after the \A4 symmetry breaking.

%
\subsection{Mass spectrum of scalars}\label{subsec:massspect}

After SSB the scalars acquire masses which  can be computed using the scalar potential \eqref{eq:scalarpot} together with the Eqs.~\eqref{eq:vev} and~\eqref{eq:doubletexp}.  To simplify our expressions for masses of the scalars, we define the following combinations of couplings:
\begin{align} \label{eq:coupl}
&\Lambda=\lambda_2 +\lambda_3 + 2\lambda_4+\lambda_5 \, ,  \nonumber \\
&\kappa_1=(-3\lambda_3-6\lambda_4+2\lambda_6+\lambda_7+\lambda_8) \, , \nonumber \\ 
&\kappa_2=(-3\lambda_3-2\lambda_4-4\lambda_5-2\lambda_6+\lambda_7+\lambda_8) \, , \nonumber \\
&\kappa_3=(-3\lambda_3-4\lambda_4-2\lambda_5+\lambda_8) \, , \nonumber \\
&\alpha=\lambda_9+\lambda_{10}+2\lambda_{11}, \nonumber \\
&\zeta=\lambda_{12}+\lambda_{13}+\lambda_{14}+\lambda_{15} \, .
\end{align}

The  mass matrix for the CP even electrically neutral scalars in the basis $\left( \phi^{ 0}, \eta_1^{ R},  \eta_{2}^{ R}, \eta_{3}^{ R}\right)$ and  $\left( \phi^{ 0}, \eta_1^{ R},  \eta_{2}^{ R}, \eta_{3}^{ R}\right)^T$  has a block diagonal form given as follows:
\begin{align}
\mathcal{M}^R_{neutral}=\left(
\begin{array}{cc}
\mathcal{M}_{H_1H_2}&0\\
0&\mathcal{M}_{{\eta}^{R}_2 {\eta}^{R}_3}\\
\end{array}
\right) \,. 
\end{align}
Note that the block diagonal form of $\mathcal{M}^R_{neutral}$ is a reflection of the unbroken residual \Z2 symmetry. Recall that $\Phi$ and $\eta_1$ are \Z2 even while $\eta_2$ and $\eta_3$ are \Z2 odd. Since the \Z2 symmetry remains unbroken, the even particles are mixing with each other, and odd particles mix among themselves leading to this peculiar block diagonal form. The matrices $\mathcal{M}_{H_1H_2}$ and $\mathcal{M}_{{\eta}^{R}_2 {\eta}^{R}_3}$ are given by:
\begin{align} \label{eq:massmatcpeven}
\mathcal{M}_{H_1H_2} & =\left(
\begin{array}{cc}
2\lambda_1 v_1^2 & \alpha v_1 v_2\\
 \alpha v_1 v_2&2\Lambda v_2^2\\
\end{array}
\right),\quad 
\mathcal{M}_{{\eta}^{R}_2 {\eta}^{R}_3} =\frac{1}{2}\left(
\begin{array}{cc}
\kappa_1 v_2^2 & 3 \zeta v_1 v_2 \\
 3 \zeta v_1 v_2 &\kappa_1 v_2^2\\
\end{array}
\right)
\;.
\end{align}
%
The mass matrix for CP odd electrically neutral scalars in the basis
$\left(  \sigma_1^{ 0}, \eta_1^{ I}, \eta_{2}^{ I}, \eta_{3}^{ I}\right)$ and  $\left(  \sigma_1^{ 0}, \eta_1^{ I}, \eta_{2}^{ I}, \eta_{3}^{ I}\right)^T$ as well as the charged scalars mass matrix in the basis $\left(\phi^{-},\eta_1^{-},\eta_2^{-},\eta_3^{-}\right)$ and $\left(\phi^{+},\eta_1^{+},\eta_2^{+},\eta_3^{+}\right)^T$ are expressed as:
\begin{align}
\mathcal{M}^I_{neutral}=\left(
\begin{array}{cc}
\mathcal{M}_{GA}&0\\
0&\mathcal{M}_{{\eta}^{I}_2 {\eta}^{I}_3}\\
\end{array}
\right), \quad \mathcal{M}_{charged}=\left(
\begin{array}{cc}
\mathcal{M}_{G^{\pm}H^{\pm}}&0\\
0&\mathcal{M}_{{{\eta}^{{\pm}}_2 {\eta}^{{\pm}}_3}}\\
\end{array}
\right). 
\end{align}
Again the block diagonal form of these matrices reflects their transformation under the unbroken \Z2 symmetry.
The four matrices $\mathcal{M}_{GA}$, $\mathcal{M}_{{\eta}^{I}_2 {\eta}^{I}_3}$, $\mathcal{M}_{G^{\pm}H^{\pm}}$ and $\mathcal{M}_{{\eta}^{\pm}_2 {\eta}^{\pm}_3}$
are given as:
\begin{align} \label{}
\mathcal{M}_{GA} &=-2\lambda_{11}\left(
\begin{array}{cc}
 v_2^2&- v_1 v_2\\
-v_1 v_2&v_1^2\\
\end{array}
\right), \quad 
\mathcal{M}_{{\eta}^{I}_2 {\eta}^{I}_3}=\frac{1}{2}\left(
\begin{array}{cc}
 \kappa_2 v^2_2 - 4 \lambda_{11} v^2_1&\zeta v_1 v_2\\
\zeta v_1 v_2 &\kappa_2 v^2_2 -4 \lambda_{11} v^2_1\\
\end{array}
\right), 
\nonumber \\
\mathcal{M}_{G^{\pm}H^{\pm}} & = -\frac{(\lambda_{10}+2\lambda_{11})}{2} \left(
\begin{array}{cc}
 v_2^2&- v_1 v_2\\
- v_1 v_2&v_1^2\\
\end{array}
\right), \nonumber \\
\mathcal{M}_{{\eta}^{\pm}_2 {\eta}^{\pm}_3} & =\frac{1}{2}\left(
\begin{array}{cc}
\kappa_3 v^2_2 -(\lambda_{10} + 2 \lambda_{11}) v^2_1&\zeta v_1 v_2\\
\zeta v_1 v_2 &\kappa_3 v^2_2 -(\lambda_{10} + 2 \lambda_{11}) v^2_1\\
\end{array}
\right).
\end{align}
The matrices $\mathcal{M}_{GA}$ and $\mathcal{M}_{G^{\pm}H^{\pm}}$ both have one vanishing eigenvalue (i.e., these are rank-1 matrices) corresponding to the neutral and charged Goldstone bosons corresponding to the $Z$ and $W^{\pm}$ gauge bosons, respectively. After diagonalization, the mass spectrum of the physical particles becomes:
 \begin{align}
 \label{eq:mass_eq}
 m_{H_1, H_2}^2 & =\lambda_1 v_1^2 +\Lambda v_2^2 \mp\sqrt{(\lambda_1 v_1^2 + \Lambda v_2^2 )^2 + v_1^2 v_2^2 (\alpha^2-4 \Lambda \lambda_1)} \, , \nonumber \\
  m_{A}^2 & =-2  \lambda_{11}\left( v_1^2 + v_2^2\right), ~~
  m_{G}^2  =0 \, , \nonumber \\
  m^2_{H^{\pm}} & =-(\lambda_{10}+2\lambda_{11}) \left( v_1^2 + v_2^2\right)/2, ~~
  m_{G^{\pm}}^2  =0 \, , \nonumber \\
 {m^2_{{\eta}^{R}_2}} & =  (\kappa_1 v_2^2 - 3\zeta v_1 v_2 )/2,~~ 
  {m^2_{{\eta}^{I}_2}}  =(\kappa_2 v_2^2 - 4\lambda_{11} v_1^2 - \zeta v_1 v_2  )/2 \, ,\nonumber \\
 {m^2_{{\eta}^{R}_3}} & ={m^2_{{\eta}^{R}_2}}+3 \zeta v_1 v_2, ~~
 {m^2_{{\eta}^{I}_3}}  ={m^2_{{\eta}^{I}_2}}+\zeta v_1 v_2 \, , \nonumber \\
 {m^2_{{\eta}^{\pm}_2}} & = (\kappa_3 v_2^2 - (\lambda_{10}+2\lambda_{11})v_1^2 - \zeta v_1 v_2 )/2,~~
 {m^2_{{\eta}^{\pm}_3}}  ={m^2_{{\eta}^{\pm}_2}}+\zeta v_1 v_2 \, .
\end{align}
We identify $H_1$ as the SM like Higgs and take its mass $125.25 \pm 0.17$ GeV~\cite{ParticleDataGroup:2022pth} throughout the whole discussion. From Eqs.~\eqref{eq:mass_eq} and \eqref{eq:coupl}, it becomes evident that the masses of the scalars are exclusively determined by the VEVs and the quartic $\lambda$ couplings. Since the $\lambda$s are constrained by the perturbativity condition Eq.~\eqref{eq:perturbcn}, it imposes a theoretical upper limit $m_{\rm{scalar}} \approx \lambda_i v \approx \sqrt{4 \pi} \times 246/\sqrt{2}\,  \rm{GeV} \lsim 600$ GeV on all scalar masses\footnote{This limit exists in our model because we have broken the $A_4$ symmetry only spontaneously. The addition of soft $A_4$ symmetry breaking terms in the potential can allow us to have scalar and DM masses larger that this limit~\cite{Pramanick:2017wry}.} \cite{Boucenna:2011tj}. 
 This implies that the DM mass in our model has an upper bound $\lesssim 600$ GeV \footnote{Even if DM is fermionic, this upper limit on its mass still applies as the DM has to the lightest particle in the dark sector. Since all \Z2 odd scalars have their masses $\lesssim 600$ GeV, the fermionic DM should also be $\lesssim 600$ GeV.}. This distinctive feature of our model is a direct consequence of the \A4 flavor symmetric approach, which distinguishes it from other scotogenic as well as (inert) Two-Higgs-Doublet models.
\subsection{Neutrino sector}\label{sec:NuSector}
Having discussed the spontaneous \A4 $\to$ \Z2 breaking and the scalar mass spectrum, we now turn to the generation of neutrino masses. According to the charge assignment outlined in Table~\ref{tab:FieldCharge}, the $SU(3)_C \otimes SU(2)_L \otimes U(1)_Y \otimes A_4$ invariant Yukawa Lagrangian that describes the leptonic sector can be expressed as follows:
\begin{eqnarray}
\label{eq:YukawaLag}
-\mathcal{L}_y &=& y_{11}(\bar{L}_1)_1\Phi (e_{R_1})_1 +  y_{22}(\bar{L}_2)_{1''}\Phi (e_{R_2})_{1'} + y_{33}(\bar{L}_3)_{1'}\Phi (e_{R_3})_{1''} + y_1 (\bar{L}_1)_1 \left(\tilde{\eta} N \right)_1 \nonumber \\
&+& y_2 (\bar{L}_2)_{1''} \left(\tilde{\eta} N \right)_{1'} + y_3 (\bar{L}_3)_{1'} \left(\tilde{\eta} N \right)_{1''} + M \left(\bar{N}^c N\right)_1 + h.c. \;, 
\end{eqnarray}
where subscript $(\cdots)_{1/1'/1''}$ denotes the \A4 transformation properties of the fields enclosed inside the parentheses. In terms of their components, the above Lagrangian can be written as follows:
 \begin{align}
- \mathcal{L}_y &= y_{11}(\bar{L}_1)_1\Phi (e_{R_1})_1 +  y_{22}(\bar{L}_2)_{1''}\Phi (e_{R_2})_{1'} + y_{33}(\bar{L}_3)_{1'}\Phi (e_{R_3})_{1''} \nonumber \\ 
&+ y_1(\bar{L}_1)_1\left(\begin{pmatrix}
\tilde{\eta}_1 \\
\tilde{\eta}_2 \\
\tilde{\eta}_3 \\
\end{pmatrix}_3  \begin{pmatrix}
N_1 \\
N_2 \\
N_3 \\
\end{pmatrix}_3 \right)_1  
+y_2(\bar{L}_2)_{1''} \left(\begin{pmatrix}
\tilde{\eta}_1 \\
\tilde{\eta}_2 \\
\tilde{\eta}_3 \\
\end{pmatrix}_3  \begin{pmatrix}
N_1 \\
N_2 \\
N_3 \\
\end{pmatrix}_3 \right)_{1'}\nonumber \\ 
&+ y_3(\bar{L}_3)_{1'}\left(\begin{pmatrix}
\tilde{\eta}_1 \\
\tilde{\eta}_2 \\
\tilde{\eta}_3 \\
\end{pmatrix}_3  \begin{pmatrix}
N_1 \\
N_2 \\
N_3 \\
\end{pmatrix}_3 \right)_{1''} + M\left(\begin{pmatrix}
\bar{N}^c_1 \\
\bar{N}^c_2 \\
\bar{N}^c_3 \\
\end{pmatrix}_3  \begin{pmatrix}
N_1 \\
N_2 \\
N_3 \\
\end{pmatrix}_3 \right)_1
+ h.c. \;, 
\end{align}
which using \A4 multiplication rules of Appendix~\ref{sec:app-A4} can be further expanded as:
\begin{align}\label{eq:YukawaLag}
- \mathcal{L}_y & = y_{11}\bar{L}_1\Phi e_{R_1} +  y_{22}\bar{L}_2\Phi e_{R_2} + y_{33}\bar{L}_3\Phi e_{R_3}  + y_1 \bar{L}_1 \left(\tilde{\eta}_1 N_1 + \tilde{\eta}_2 N_2 +\tilde{\eta}_3 N_3  \right) \nonumber \\ &+ y_2 \bar{L}_2 \left(\tilde{\eta}_1 N_1 + \omega \tilde{\eta}_2 N_2 + \omega^2 \tilde{\eta}_3 N_3  \right) + y_3 \bar{L}_3 \left(\tilde{\eta}_1 N_1 + \omega^2 \tilde{\eta}_2 N_2 + \omega \tilde{\eta}_3 N_3  \right) \nonumber \\ &+ M(\bar{N}^c_1 N_1 + \bar{N}^c_2 N_2 + \bar{N}^c_3 N_3) +  h.c. \;.
\end{align}
where $\tilde{\eta}=i \tau_2 \eta^{\ast}$; $\tau_2$ is the second Pauli matrix and $y_{ii}$, $y_i$ for $i = 1, 2, 3$ are the Yukawa couplings, $\omega$ is the cubic root of unity and $M$ is the Majorana mass of  fermion $N$. 
\begin{figure}[b]
\centering
\includegraphics[width=0.9\textwidth]{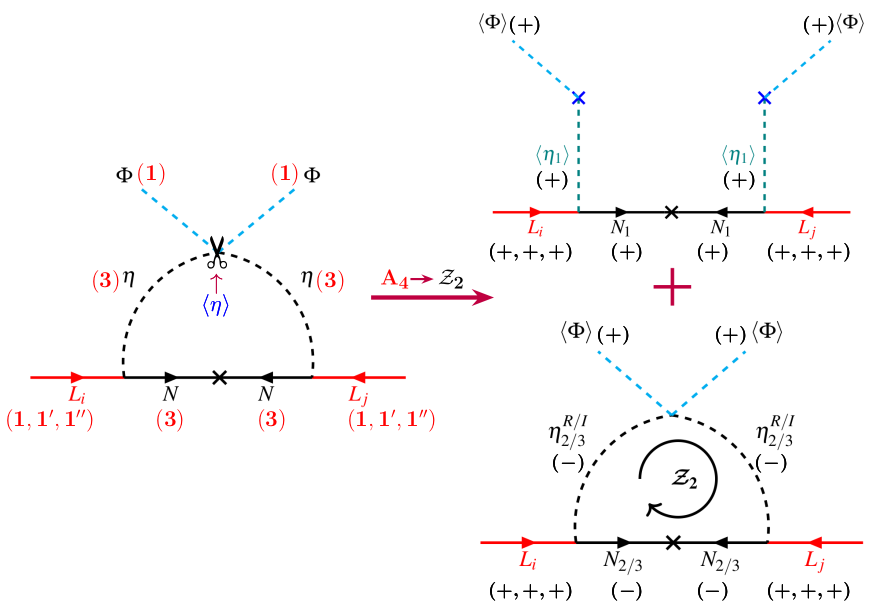}
\caption{\footnotesize 
Schematic diagram showing how \A4 $\to$ \Z2 cuts the scotogenic loop leading to a hybrid scoto-seesaw mass mechanism. The left panel is flavor $A_4$ invariant with both particles ($\eta, N$) running inside the loop being triplets of the $A_4$ group. The right panel is invariant under the residual \Z2 subgroup which also serves as the dark symmetry. The $A_4$ symmetry is broken spontaneously to a residual \Z2 subgroup by the VEV of the scalar $\eta$, thus cutting the radiative loop. This results in a scoto-seesaw mass mechanism i.e., a type-I seesaw and a one-loop level radiative model as shown in the right panel. The residual \Z2 subgroup running inside the loop serves as the dark symmetry. Before (after) symmetry breaking \A4 (\Z2) charges have been shown in the left (right) panel of the diagram. }
\label{fig:breakingA4}
\end{figure}

As discussed in Sec.~\ref{sec:scalarsec} the \A4 symmetry is broken to the residual \Z2 subgroup by the VEV of the first component, $\eta_1$ of the triplet $\eta$, while the other two components $\eta_2$ and $\eta_3$ do not acquire VEV. Since, $N$ is also a triplet of \A4, after the \A4 symmetry is broken, its components transform as
\begin{equation}
    N_1 \to + N_1 \, ,  \quad  N_2 \to - N_2  , \quad  N_3 \to - N_3 .
\end{equation}
under the residual \Z2 subgroup. The lepton doublets ($L_i$) and charged lepton singlets ($e_{R_i}$); $i = 1,2,3$ being \A4 singlets, will be even under the residual \Z2 subgroup.  
Hence the charged leptons get masses through the standard Higgs mechanism and they can also mix among themselves. However, due to the specific charge assignment of charged leptons under \A4, the charged lepton mass matrix is diagonal and is written as $M_l = v_1 \, {\rm diag}\, (y_{11}, y_{22}, y_{33})/\sqrt{2}$.
Therefore, the observed oscillation and mixing pattern of leptons arises from the neutrino sector only.  

Coming to the neutrino mass generation, when the \A4 symmetry is unbroken, one can have a radiative loop as shown in the left panel of Fig.~\ref{fig:breakingA4}. 
%
However, since $\langle \eta_1 \rangle = \frac{v_2}{\sqrt{2}} \neq 0$, the spontaneous breaking of \A4 $\to$ \Z2 leads to ``cutting" of this radiative loop. This breaking pattern leads to the hybrid scoto-seesaw mass generation where neutrinos get mass both from a type-I seesaw as well as from the scotogenic loop as shown in the right panel of Fig.~\ref{fig:breakingA4}. This happens because after SSB the $\eta_1$ field being \Z2 even mixes with $\Phi$ and results in the type-I like seesaw diagram where the \Z2 even fermion $N_1$ acts as an intermediate connection as shown in the upper part of the right panel of Fig.~\ref{fig:breakingA4}. 
The \Z2 odd fermions $N_{2,3}$ and scalars $\eta_{2,3}$ now belong to the dark sector and they together run inside a radiative loop as depicted in the lower right panel of Fig.~\ref{fig:breakingA4}. The conserved \Z2 symmetry then ensures that the lightest particle running inside the loop can be a good DM candidate.

At the tree-level the mass matrix in the basis of $(\bar{\nu_1}^c,\bar{\nu_2}^c,\bar{\nu_3}^c, \bar{N_1}^c,\bar{N_2}^c,\bar{N_3}^c)$ and  \\ $(\nu_1,\nu_2,\nu_3,N_1,N_2,N_3)^T$ is given by
\begin{equation}
\mathcal{M}_{\nu}= \frac{1}{\sqrt{2}}
\begin{pmatrix}
0 & 0 & 0 & y_1 v_2 &0 &0\\
0 & 0 & 0 &y_2 v_2 & 0 &0 \\
0 & 0 & 0 &y_3 v_2&0 &0\\
y_1 v_2 & y_2 v_2 & y_3 v_2 &\sqrt{2} M &0 &0\\
0 & 0 & 0 &0 &\sqrt{2} M &0\\
0 & 0 & 0 &0 &0 &\sqrt{2} M\\
 \end{pmatrix} \;.
 \label{eq:big-matrix}
\end{equation}
Note that the zeros in off-diagonal terms in $\mathcal{M}_{\nu}$ are due to the conserved \Z2 symmetry which forbids coupling between even ($\nu_i, N_1$) and odd ($N_2, N_3$) fields. The mass matrix in Eq.~\eqref{eq:big-matrix}
can be further blocked diagonalized in the type-I seesaw limit.  The resulting light neutrino mass matrix can be expressed as
\begin{equation}
-  m^{(1)}_{\nu} =  m_D \mathcal{M}^{-1} m_D^T = 
\frac{v_2^2}{2 M} 
\begin{pmatrix}
y_1^2 & y_1y_2 & y_1y_3 \\
\ast & y_2^2 & y_2y_3  \\
\ast & \ast & y_3^2\\
 \end{pmatrix} \;.
\end{equation}
\begin{align}
\text{where,} \quad
m_D=\frac{v_2}{\sqrt{2}}
\begin{pmatrix}
y_1 & 0 & 0\\
y_2 & 0 & 0\\
y_3 & 0 & 0
\end{pmatrix},\,\, \quad
\mathcal{M}= {\rm diag}\, M (1, 1, 1)\, .
\end{align}
Note that just like in canonical type-I seesaw, the mass of fermion $N_1$ also gets a small seesaw correction and will change slightly from the $M$ value, whereas the tree-level masses of the other two fermions $N_2$ and $N_3$ will remain $M$  because of $A_4$ symmetry. Eventually, the degeneracy in $N_2$, $N_3$ masses is also slightly lifted due to the loop corrections.

Coming back to the light neutrino mass matrix $m^{(1)}_{\nu}$, one can immediately notice that the rank of $ m^{(1)}_{\nu} $ is one and hence only one of the neutrinos gets mass through this type-I seesaw diagram.  
However, the masses of neutrinos also get contribution from the one-loop diagram as shown by Fig.~\ref{fig:breakingA4}.
The additional mass matrix  arising from the  one-loop is given by~\cite{Ma:2006km}
\begin{align}
\label{}
(\mathcal{M}_{\nu})_{ij} =\sum_{l,k=1}^{3}\frac{Y_{ik}Y_{jk}}{32\pi^2}M_k\left[\ \frac{{(m^{R}_{\eta_l})}^2}{{(m^{R}_{\eta_l})}^2 - M^{2}_k}\ln\left(\frac{{m^{R}_{\eta_l}}}{M_k}\right)^2  - \frac{{(m^{I}_{\eta_l})}^2}{{(m^{I}_{\eta_l})}^2 - M^{2}_k}\ln\left(\frac{{m^{I}_{\eta_l}}}{M_k}\right)^2 \right]\;.
\end{align}
Because of \A4 symmetry in our model, we have $l=k$. Thus the mass matrix is simplified to
\begin{align}
\label{loop mass}
(\mathcal{M}_{\nu})_{ij}=  \sum_{k=1}^{3}Y_{ik}Y_{jk}c_k \;, 
\end{align}
where, $Y_{ik}$ and $Y_{jk}$ are Yukawa couplings at one-loop level, and 
\begin{align}
\label{c expression}
c_k = \frac{M_k}{32\pi^2}\left[\ \frac{{(m^{R}_{\eta_k})^2}}{{(m^{R}_{\eta_k})}^2 - M^{2}_k}\ln\left(\frac{{m^{R}_{\eta_k}}}{M_k}\right)^2  - \frac{{(m^{I}_{\eta_k})}^2}{{(m^{I}_{\eta_k})}^2 - M^{2}_k}\ln\left(\frac{{m^{I}_{\eta_k}}}{M_k}\right)^2 \right]\;.
\end{align}
The couplings  $Y_{ik}$ and $Y_{jk}$ are given by
\begin{align}
&Y_{11}=  y_1,  \quad   Y_{12}= y_1,  \quad   \hspace{0.3cm}    Y_{13}= y_1 \, ,      \nonumber \\
&Y_{21}=   y_2,  \quad    Y_{22}= \omega y_2,  \quad  \hspace{0.1cm}    Y_{23}= \omega^2 y_2 \, ,    \nonumber \\
&Y_{31}=   y_3 ,  \quad   Y_{32}= \omega^2 y_3,  \quad      Y_{33}= \omega y_3 \, . 
\label{eq:def-yuk}
\end{align}
Substituting \eqref{eq:def-yuk} in \eqref{loop mass} and simplifying, we get 
\begin{align}
&(\mathcal{M}_{\nu})_{11}=  y^2_1\left( c_1 + c_2 + c_3 \right),  \quad  (\mathcal{M}_{\nu})_{12}= y_1y_2\left( c_1 + \omega c_2 + \omega^2 c_3\right), \quad (\mathcal{M}_{\nu})_{13}= y_1y_3\left( c_1 + \omega^2c_2 + \omega c_3\right) \, ,
\nonumber \\
&(\mathcal{M}_{\nu})_{21}=   y_2y_1\left( c_1 + \omega c_2 + \omega^2c_3\right),  \quad    (\mathcal{M}_{\nu})_{22}=  y^2_2\left( c_1 + \omega^2c_2 + \omega c_3\right),  \quad      (\mathcal{M}_{\nu})_{23}=  y_2y_3\left( c_1 + c_2 + c_3\right) \, , 
\nonumber \\
&(\mathcal{M}_{\nu})_{31}=   y_3y_1\left( c_1 + \omega^2c_2 + \omega c_3\right),  \quad  (\mathcal{M}_{\nu})_{32}= y_3y_2\left( c_1 + c_2 + c_3\right),  \quad      (\mathcal{M}_{\nu})_{33}= y^2_3\left( c_1 + \omega c_2 + \omega^2c_3\right) \, . 
\end{align}
Thus, the light neutrino mass matrix arising from the one-loop level is given by:
\begin{center}
\begin{equation}
m^{(2)}_{\nu} =
\begin{pmatrix}
y^2_1d_1 & y_1y_2d_2 & y_1y_3d_3  \\
\ast & y^2_2d_3 & y_2y_3d_1 \\
\ast & \ast & y^2_3d_2 \\
\end{pmatrix} \;.
\end{equation}
\end{center}
%
\begin{align}\label{eq:dvalues}
\text{where,} \quad \quad d_1  = c_1+c_2+c_3, \quad d_2 =  c_1+\omega c_2+\omega^2c_3, \quad d_3 =  c_1+\omega^2 c_2+\omega c_3.
\end{align}
Note that here  $d_1$ is real, whereas due to the presence of $\omega$ and $\omega^2$, $d_2$ and $d_3$ are complex conjugates of each other. The expressions of  $c_1$, $c_2$, $c_3$ can be derived from Eq.~\eqref{c expression}. It is to be noted here that as all the Yukawa couplings are considered real, the CP violation in the leptonic sector arises from the complex nature of $ d_2 $ and $ d_3 $ only.

Combining both the tree-level seesaw as well as one-loop level scotogenic contributions,  the ``scoto-seesaw" mass matrix for the light neutrinos can be expressed as:
\begin{equation}\label{eq:NuMassMatrix}
m^{(TOT)}_{\nu} = m^{(1)}_{\nu} + m^{(2)}_{\nu} \equiv
\begin{pmatrix}           
A & C & \tilde{C} \\
\ast & B & D \\
\ast & \ast & \tilde{B} \\                                
 \end{pmatrix} \;.
\end{equation}
where,
\begin{align} \label{eq:numatrixpara}
&A =  y^2_1\left(d_1-\frac{v_2^2}{2 M}\right),  \quad \quad D =  y_2y_3\left(d_1-\frac{v_2^2}{2 M}\right),  \nonumber \\
&B =  y^2_2\left(d_3-\frac{v_2^2}{2 M}\right), \quad \quad  \tilde{B} = y^2_3\left(d_2-\frac{v_2^2}{2 M}\right), \nonumber \\ 
&C =  y_1y_2\left(d_2-\frac{v_2^2}{2 M}\right), \quad  \tilde{C} =  y_1y_3\left(d_3-\frac{v_2^2}{2 M}\right).
\end{align}
By virtue of $d_1$ being real and  $d_2$, $d_3$ being conjugates of each other, the parameters $A$ and $D$ are real, whereas parameters $B$, $ \tilde{B}$, $C$ and $\tilde{C}$ are complex.

\section{Numerical results for neutrino sector}
\label{sec:Nuprediction}

In this section we aim to discuss our main results and predictions for the neutrino sector. To begin with, note that, in the scoto-seesaw models the Yukawa couplings appearing in the seesaw part and those appearing in the scotogenic part are independent of each other~\cite{Rojas:2018wym,Mandal:2021yph}. Thus, in general, the typical scoto-seesaw models 
allow for two types of limiting cases depending upon the fermion masses:
\begin{itemize}
\item \textbf{Case-I:} Scalar masses $m_{\eta_i}$ in GeV-TeV range, fermion masses, $M_i\sim m_{\eta_i}$.
\item \textbf{Case-II:} Scalar masses $m_{\eta_i}$ in GeV-TeV range, fermion masses, $M_i >> m_{\eta_i}$. 
\end{itemize}
From the purely aesthetic point of view, the case-I is less appealing as it necessarily requires smaller Yuakwa couplings in the $\sim 10^{-8} - 10^{-6}$ range while case-II allows for Yukawa couplings up to $\mathcal{O} (1)$. However, both limits are perfectly allowed by the canonical scoto-seesaw model and its variants.

In our case, due to the \A4 symmetry, the same Yukawa couplings (see~\eqref{eq:YukawaLag} and discussion in Sec.~\ref{sec:NuSector}) appear in both seesaw and scotogenic parts. Thus our model is more constrained than typical scoto-seesaw models. 
In fact, during the numerical analysis, we found that if the masses of the fermions ($N_1, N_2, N_3$) are $\leq 10^5$ GeV, we can not simultaneously satisfy the two mass-squared differences $\Delta m^2_{\rm {atm}}$ and $\Delta m^2_{\rm {sol}}$ of neutrino oscillations within their current 3$\sigma$ range, see Appendix~\ref{app:fermionicDM} for more details. Therefore, in our model, case-I is automatically ruled out as only higher values of fermion masses $m_{N_i} >> m_{\eta_i}$ can fit the observed neutrino oscillation constraints. This also implies that the fermionic dark matter case where $m_{N_i} < m_{\eta_i} \lesssim 600$ GeV
is also ruled out as discussed in Appendix~\ref{app:fermionicDM}.

To perform the numerical analysis the input parameters are varied following Table~\ref{tab:scan1}. 
\begin{table}[h!]
\centering
\begin{tabular}{|c|c|c|c|}
\hline
Parameters & Range &Parameters & Range \\
\hline
$M_{1,2,3}$ (in GeV) & $[10^6, 10^{12}]$ &$y_i$ &$[10^{-6}, 10^{-2}]$ \\
\hline
$\lambda_1$ & $[10^{-3}, \sqrt{4\pi}]$ & $|\lambda_{2,3...,10}|$ & $[10^{-6}, \sqrt{4\pi}]$  \\
$\lambda_{11}$ & $-[10^{-6}, \sqrt{4\pi}]$ & $\lambda_{12,13,14,15}$ & $[10^{-8}, 10^{-2}]$ \\
\hline
\end{tabular}
\caption{\footnotesize \centering Value range for the numerical parameter scan in the neutrino sector.}
 \label{tab:scan1}
\end{table} 
The value of Higgs mass has been fixed within its experimental 3$\sigma$ range $125.25 \pm 0.17$ GeV~\cite{ParticleDataGroup:2022pth}, whereas all other scalars masses have been varied up to 600 GeV, keeping $\eta_2^R$ as the lightest dark sector particle, hence a good DM candidate. 
While the neutrino sector constraints primarily depend on the masses of $N_i$ and the Yukawa couplings, the $\lambda$s will play a key role in the dark sector analysis, see Sec.~\ref{sec:DarkSector}.

After performing the numerical analysis we find that our results are in good agreement with the three mixing angles ($\theta_{12}$, $\theta_{13}$, and  $\theta_{23}$) as well as two mass-squared differences ($\Delta m^2_{\rm {atm}}$ and $\Delta m^2_{\rm {sol}}$) of neutrino oscillation data~\cite{Capozzi:2021fjo, deSalas:2020pgw, Esteban:2020cvm} only for normal ordering (NO) of neutrino masses. Therefore all our results are obtained using NO of neutrino masses. 

\subsection{Predictions for CP phases}

\begin{figure}[h!]
\centering
\includegraphics[width=0.44\textwidth]{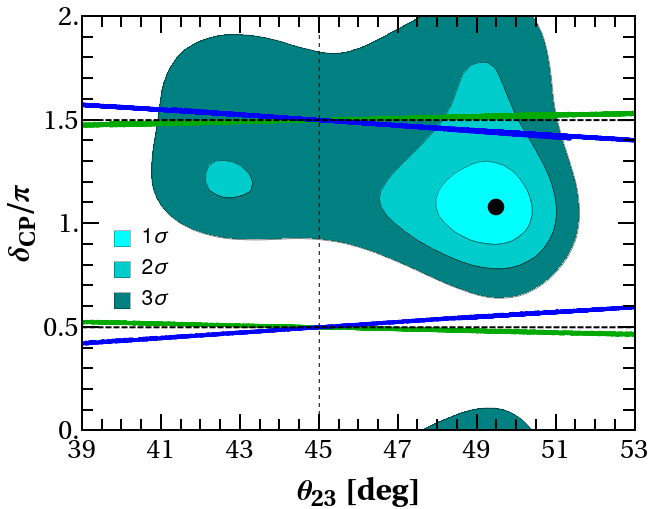}
\hspace{1.0cm}
\includegraphics[width=0.42\textwidth]{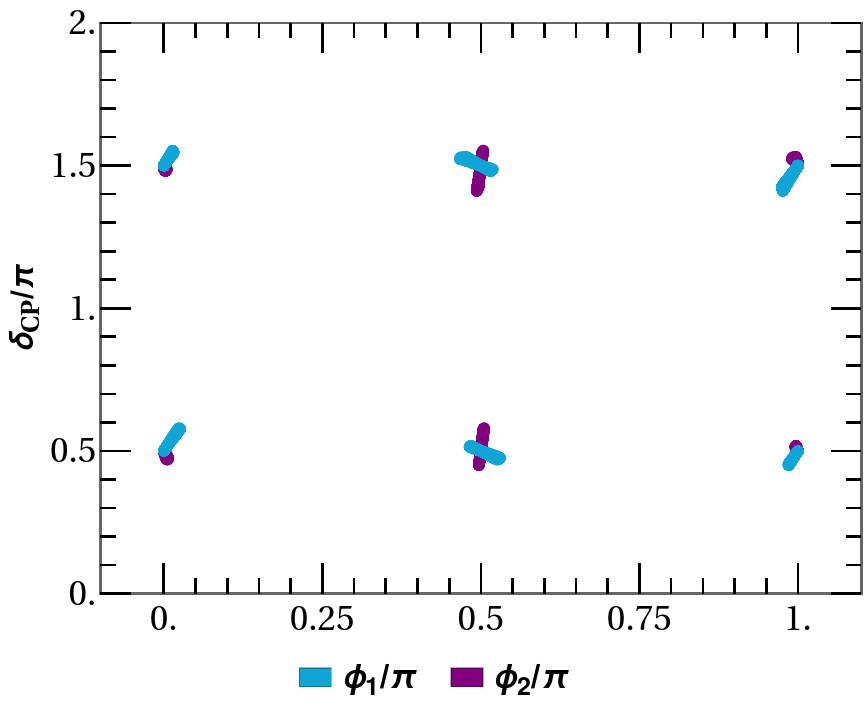}
\caption{\footnotesize \textbf{Left panel}: Predictions of the model in the  ($ \theta_{23} -\delta_{\rm{CP}} $) plane are shown by the green and blue points. The cyan-colored contours correspond to the current $1,2,3 \, \sigma $ confidence level of the latest global-fit data~\cite{deSalas:2020pgw}.  The intersection of dotted-black lines corresponds to the exact $ \mu-\tau $ reflection symmetry limit.  
\textbf{Right panel}: Correlation between Dirac ($\delta_{\rm{CP}}$) vs Majorana CP phases ($\phi_1$, $\phi_2$) are shown.
}
\label{fig:NuSector1}
\end{figure}

In this section we discuss how \A4 symmetry leads to predictions regarding the flavor structure of neutrinos. We start with observing that the mass matrix in Eq.~\eqref{eq:NuMassMatrix}, for $ \tilde{B} = B^* $ and $ \tilde{C} = C^* $, becomes $ \mu-\tau $ reflection symmetric~\cite{Harrison:2002et}. In this limit  the atmospheric mixing angle, $ \theta_{23} = 45^\circ$ and Dirac CP phase $ \delta_{\rm{CP}} = \pi/2 $ or $ 3\pi/2 $. 
From Eq.~\eqref{eq:numatrixpara} we see that since $d_2$ and $d_3$ are complex conjugate to each other, thus
the condition for the  $ \mu-\tau $ reflection symmetry i.e. $ \tilde{B} = B^* $ and $ \tilde{C} = C^* $ is obtained simply when  the Yuakwa couplings $y_2 = y_3$. 
The exact $ \mu-\tau $ reflection symmetry limit is shown by the intersection of dotted-black lines in the left panel of Fig.~\ref{fig:NuSector1}. 

In general, the two Yukawa couplings $y_2$ and $y_3$ are independent parameters and can have different values. Whenever $y_2 \neq y_3$ in the neutrino mass matrix (see Eq.~\eqref{eq:NuMassMatrix}), we depart from the exact $ \mu-\tau $ reflection symmetry. Thus our model leads to the ``generalized $ \mu-\tau $ reflection symmetry" as shown by the green and blue points in the left panel of Fig.~\ref{fig:NuSector1}.
Depending upon the choice of $y_2$ and $y_3$, the mixing angle $\theta_{23}$ will shift to lower or upper octant as follows:
\begin{align}\label{Eq:Octant}
&y_2 < y_3 \Rightarrow  \theta_{23} < 45^{\circ} \,  ,\nonumber \\
\quad &y_2 > y_3 \Rightarrow  \theta_{23} > 45^{\circ} \, .
\end{align}
%
Note that in Fig.~\ref{fig:NuSector1} both green and blue points lie very close to maximal $ \delta_{\rm{CP}}$ values predicting large CP violation in the leptonic sector. Furthermore, they are symmetric with respect to reflection around $ \delta_{\rm{CP}} = \pi$ value.
The  gaps between the two sets of points
correspond to points that fail to either satisfy the $3 \sigma$ constraints of the other two mixing angles $\theta_{12}$ and $\theta_{13}$ or the two mass-squared differences, leading to the splitting between the green and blue points.
For comparison, we also show the latest global analysis of neutrino oscillation data \cite{deSalas:2020pgw} using the cyan color contours.
It can be inferred that only the solutions around $ \delta_{\rm{CP}} =  1.5\, \pi $ are consistent with the latest neutrino data at $3 \sigma $ significance level.
This can serve as an important test of our model as an improved measurement of $ \delta_{\rm{CP}}$ significantly away from 
$ 1.5\, \pi $ value cannot be explained by our model. 
Indeed future experiments like DUNE~\cite{DUNE:2015lol} can measure $ \delta_{\rm{CP}}$ with much higher accuracy and can confirm or rule out our model entirely.  

In the right panel, we have shown the correlation of the Dirac CP phase $\delta_{\rm{CP}}$ with the Majorana CP phases $\phi_1$ and $\phi_2$. From here, it is evident that both the Majorana CP phases are highly constrained and lie close to their CP-conserving values. This pattern can also be understood by recalling that in the exact $\mu-\tau$ refection symmetry limit, the Majorana phases can have only $0$ or $\frac{\pi}{2}$ values. Since our model has generalized $\mu - \tau$ reflection symmetry, the phases deviate by a small amount from their exact $\mu - \tau$ reflection symmetry limit.
The importance of these highly constrained Majorana phases will be analyzed for $0\nu\beta\beta$ decay. 

\subsection{Lower limit on the lightest neutrino mass}
\begin{figure}[h!]
\centering
\includegraphics[width=0.50\textwidth]{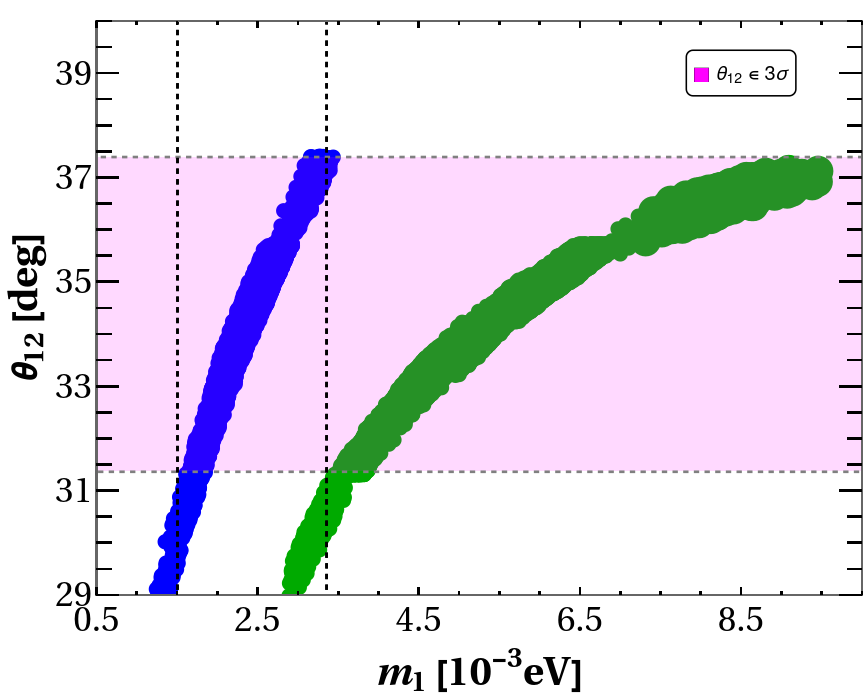}
\caption{\footnotesize \centering Correlation between  mixing angle $\theta_{12}$ and the lightest neutrino mass $m_{\rm{lightest}} = m_1$. }
\label{fig:mlight}
\end{figure}

In our model, we also have the prediction for the lightest neutrino mass $m_{\rm{lightest}} = m_1$, shown in Fig.~\ref{fig:mlight}.

 To illustrate that, we present a correlation plot between $m_1$ and mixing angle $\theta_{12}$ using the green and blue points. The reason behind these two sets remains the same as the left panel of Fig.~\ref{fig:NuSector1}.
From Fig.~\ref{fig:mlight}, one can see that once we impose the $3\sigma$ range of $\theta_{12}$ from the global fit data (see the pink region), we find a small range for the $m_1$ mass. 
We find the lowest allowed values as  $m_1 \sim 1.5$ (3.35) meV from the blue (green) points. 
It is important to note that these minimum allowed values of neutrino mass are crucial for predicting the effective Majorana neutrino mass in $0\nu\beta\beta$ decay, a topic that will be discussed in the following section.

\subsection{Predictions for  $ 0\nu \beta \beta $ decay and beta decay}

Turning to other observables in the neutrino sector, one of the most important to consider is neutrinoless double beta decay ($0\nu\beta\beta$) which is a robust way for searching for lepton number violation and Majorana nature of neutrinos~\cite{Schechter:1981bd}. The half-life of $0\nu\beta\beta$ process is given by
\begin{equation}
\frac{1}{T^{0\nu}_{1/2}} = G_{0\nu}|M_{0\nu}(A,Z)|^{2} |\langle m_{ee} \rangle|^{2} \;.
\label{eq:halflifedoublebeta}
\end{equation}
where, $  G_{0\nu}$ represents the two-body phase-space factor, $ M_{0\nu} $ is the nuclear matrix element and 
$\langle m_{ee} \rangle$ is the effective Majorana neutrino mass.
%
In the standard PDG parametrization, $\langle m_{ee}\rangle$ is expressed as,
\begin{align} \label{eq:meef}
 \langle m_{ee}\rangle \equiv  \bigg | \sum_{j} U^2_{ej}m_j \bigg |= \big |c^2_{12}c^2_{13}m_1 e^{-2 \mathit{i} \phi_1}+s^2_{12}c^2_{13}m_2 e^{-2 \mathit{i} \phi_2}+s^2_{13}m_3 e^{2 \mathit{i} \delta_{\rm{CP}}} \big | \;.
\end{align}
where, $c_{ij}=\cos \theta_{ij}$, $s_{ij}=\sin \theta_{ij}$, and $m_i$ are neutrino masses, $\phi_1$, $\phi_2$ are Majorana CP phases while $\delta_{\rm{CP}}$ is the Dirac CP phase. The current limits on $0\nu\beta\beta$ decays from  KamLAND-Zen~\cite{KamLAND-Zen:2022tow}, EXO~\cite{Agostini:2017jim}, GERDA Phase-II~\cite{GERDA:2018pmc}, and CUORE~\cite{CUORE:2017tlq} experiments can be used to put constraints on $\langle m_{ee}\rangle$.

\begin{figure}[!h]
\centering
\includegraphics[width=0.45\textwidth]{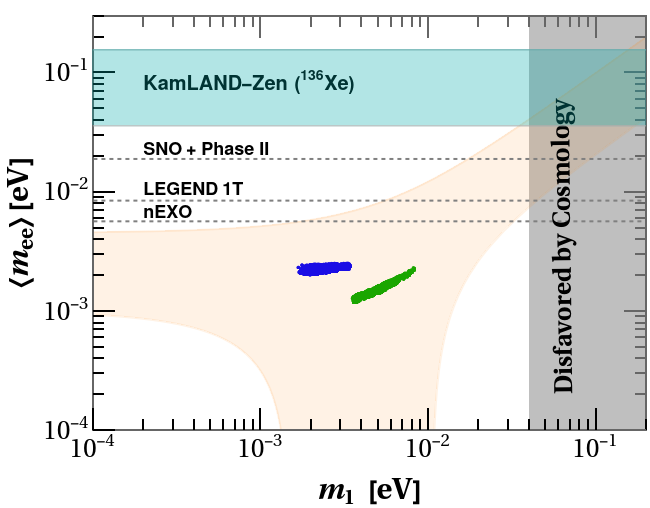}
\hspace{1.0cm}
\includegraphics[width=0.45\textwidth]{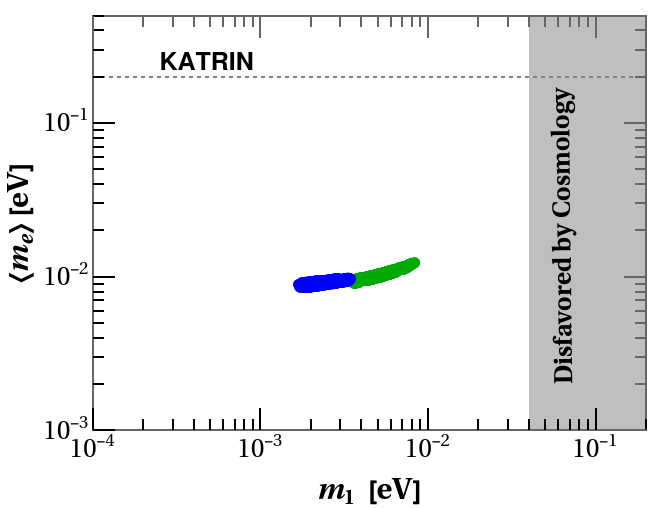}
\caption{\footnotesize \textbf{Left:} The effective Majorana neutrino mass $\langle m_{ee} \rangle$, \textbf{Right:} the effective mass of the electron neutrino $\langle m_{e} \rangle$, as a function of the lightest neutrino mass are shown. The green (blue) color region shows our model predictions whereas the orange region in the left panel shows the parameter space allowed by current oscillation data for NO. See text for more details. }
\label{fig:NuSector2}
\end{figure}

The constrained values of the CP phases and masses in our model imply that the value of $ \langle m_{ee}\rangle$ is strongly constrained as shown by the blue and green regions in the left panel of Fig.~\ref{fig:NuSector2}.
%
The color code remains the same as in the previous plots.
Again the splitting between green and blue points is due to the non-fulfillment of $3 \sigma$ constraints of either the mixing angles or the two mass-squared differences.
The orange region represents the full model-independent parameter space allowed by the latest global-fit data~\cite{deSalas:2020pgw}. The vertical gray band shows the constraint on the light neutrino mass arising from the  {\it Planck} (TT, TE, EE + lowE + lensing + BAO) dataset, which has set an upper bound on the sum of neutrino masses $\sum m^{}_{i} <0.12~{\rm eV}$~\cite{Planck:2018vyg}. 
The horizontal cyan band indicates the current experimental limits from KamLAND-Zen ($[36–156]$ meV)~\cite{KamLAND-Zen:2022tow}, while the black-dashed lines correspond to the projected sensitivities for ${\rm SNO}+$ Phase II ($[19-46]$ meV)~\cite{SNO:2015wyx}, LEGEND 1000 ($[8.5-19.4]$ meV)~\cite{LEGEND:2021bnm}, nEXO ($[5.7-17.7]$ meV)~\cite{nEXO:2017nam} at 90\% C.L..
One could see that the model predictions remain outside the testable range of the next generation $ 0\nu \beta \beta $ decay experiments. Thus, any observation of $0\nu\beta\beta$ decay by these experiments will completely rule out our model.

In the right panel of Fig.~\ref{fig:NuSector2}, we have shown our model prediction for the effective mass of the electron neutrino $\langle m^{}_{e} \rangle$, defined as
\begin{align} \label{eq:me}
 \langle m^{}_{e}\rangle \equiv \sqrt {\sum_{j}  |U_{ej} |^2 m^2_j}  = \sqrt{c^2_{12}c^2_{13}m^2_1 + s^2_{12}c^2_{13} m^2_2 +s^2_{13}m^2_3  }\;.
\end{align}
The horizontal dashed line in the right panel of Fig.~\ref{fig:NuSector2} corresponds to the expected full dataset sensitivity of KATRIN~\cite{KATRIN:2021uub} ($\langle m^{}_{e}\rangle< 0.2 $ eV) while the vertical gray line corresponds to the Planck limit. Thus a measurement of $\langle m^{}_{e} \rangle$ by KATRIN can also completely rule out our model.

\section{Dark sector}\label{sec:DarkSector}

We now come to the dark sector predictions of our model. To start with, recall that in the canonical scotogenic model~\cite{Ma:2006km}, the lightest $\mathcal{Z}_2$ odd particle (among scalar $\eta$ and fermions $N$)  running inside the loop, plays the role of a possible DM candidate. 
Depending on the masses of the dark sector particles, one ends up with two possible DM candidates: the lightest neutral dark scalar or the lightest dark fermion.

In our case, after \A4 $\to$ \Z2 breaking, the components of \A4 triplets $ \eta $ and $ N $ fields transform under the residual \Z2 symmetry as follows (see Appendix~\ref{sec:app-A4}):
\begin{equation}\label{residualZ2}
\begin{array}{lcrlcr}
N_1 &\to& +N_1\,,\quad& \eta_1 &\to& +\eta_1 \, ,\\  
N_{2,3} &\to& -N_{2,3}\,,\quad& \eta_{2,3} &\to& -\eta_{2,3} \;.
\end{array}
\end{equation}
All the remaining SM fields are even under $\mathcal{Z}_2$ as they are singlets of $A_4$. Thus the dark sector in our model consists of the \Z2 odd particles $\eta_2$, $\eta_3$, $N_2$ and $N_3$ with the lightest electrically neutral particle being the DM candidate. 

As discussed in Sec.~\ref{sec:scalarsec}, masses of scalars have an upper bound $\sim$ 600 GeV.  The lower value of fermion mass ($\lesssim 600$ GeV) is not in good agreement with the two mass-squared differences ($\Delta m^2_{\rm{atm}}$ and $\Delta m^2_{\rm{sol}}$) of neutrino oscillations, as presented in Appendix~\ref{app:fermionicDM}. Therefore, in this model, the only viable option for DM is a scalar particle. In what follows, we present a comprehensive analysis of scalar DM.

\subsection{Scalar dark matter}
In the scalar sector the $SU(2)_L$ doublets $\eta_2$ and  $\eta_3$ are odd under \Z2 and their neutral components can be a DM candidate. From Eq.~\eqref{eq:doubletexp}, it is evident that after SSB we have four neutral dark scalars, namely $\eta_{2}^R$, $\eta_{2}^I$, $\eta_{3}^R$ and $\eta_{3}^I$. 
In principle, out of these four scalars,  anyone can be the lightest and can serve as a DM candidate.  For the sake of definiteness throughout this section we take $\eta_{2}^{R}$ as a scalar DM candidate with the following condition:
\begin{align} \label{eq:dmmassconstraint}
m_{\eta^R_{2}} < m_{\eta^I_{2}}, \hspace{0.1cm} m_{\eta^R_{3}}, \hspace{0.1cm} m_{\eta^I_{3}} \; .
\end{align}
Note that taking another one of them as a DM candidate will not change the main results of our analysis.
We have performed a detailed numerical scan for the model parameters taking into account the various experimental and theoretical constraints. 
To generate the allowed points, we have imposed the following conditions:   
\begin{itemize} 
\item The tree-level vacuum stability has been imposed in accordance with Eq.~\eqref{eq:vacuumstab}.
\item The positivity of pseudo scalar $(A)$ and charged Higgs $(H^{\pm})$ mass, have been ensured by taking $\lambda_{11}<0$ and $\lambda_{10}+ 2 \lambda_{11}<0$ respectively, see Eq.~\eqref{eq:mass_eq}.
\item $\lambda_{12,13,14,15}>0$ has been chosen to ensure the $m_{\eta^R_2}<m_{\eta^R_3}$.
\item In addition, we have also imposed the condition $m_{\eta^R_2}<m_{\eta^I_2}$ (the condition $\lambda_{12,13,14,15}>0$, automatically ensures that $m_{\eta^R_2}<m_{\eta^I_3}$). 
\item Note that these constraints automatically satisfy the condition mentioned in Eq.~\eqref{eq:dmmassconstraint}. 
\item The remaining couplings are allowed to vary freely following the Eq.~\eqref{eq:mass_eq}. 
\item The fermion mass $M$ and Yukawa couplings $y_i$ are varied in the range given in Table~\ref{tab:scan1}, always ensuring that the neutrino sector observables are within their $3\sigma$ range.  
\item The input parameters for scalar sector analysis are taken in the ranges provided in Table.~\ref{tab:scan1}. 
\end{itemize}
%
\subsubsection*{\textbf{Relic density}}\label{sec:relic}

\begin{figure}[h!]
\centering
\includegraphics[width=0.7\textwidth]{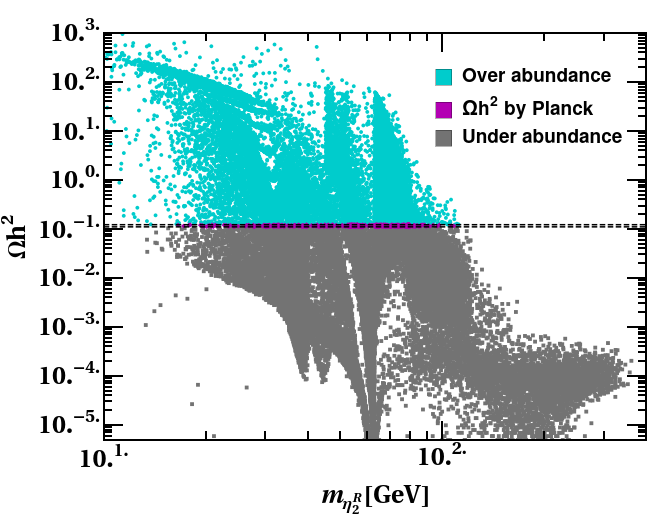}
\caption{\footnotesize  Relic density as a function of scalar DM mass $ m_{\eta_2^R} $. The cyan, magenta, and gray points depict over, observed, and under abundance of the DM relic density, respectively. 
The horizontal black dashed lines signify the latest Planck satellite data~\cite{Planck:2018vyg}.
}
\label{fig:relic}
\end{figure}

We start with the computation of the relic abundance for $\eta^R_2$ as DM. In Fig.~\ref{fig:relic}, we show our result for relic density as a function of the mass of the scalar DM $\eta^R_2$. The numerical scan is performed by varying the input parameters as given in Table~\ref{tab:scan1} and applying the constraints mentioned before. The narrow band shown by black dotted lines corresponds to the $3 \sigma$ range for the cold DM relic density reported by the Planck satellite data~\cite{Planck:2018vyg}:
\begin{align*}
0.1126 \leq \Omega_{\eta^R_2} h^{2} \leq 0.1246\;.
\end{align*}
%

%

The points that lie within this narrow band correspond to $\eta^R_2$ as $100\%$ DM candidate. These points are shown in magenta color in Fig.~\ref{fig:relic}. While the cyan/gray color points represent over/under abundance of relic density, respectively.
There are three sharp dips in Fig.~\ref{fig:relic}. The first dip is at $m_{\eta_2^R} \sim M_W/2$, the second dip is at $m_{\eta_2^R} \sim M_Z/2$ and third dip is at $m_{\eta_2^R} \sim m_{H_1}/2$, which corresponds to annihilation via s-channel W boson exchange, Z boson exchange, and Higgs boson exchange, respectively as shown in Fig.~\ref{fig:scanni1} in Appendix~\ref{feyndiag1}. Also, one can notice that the dip at $m_{\eta_2^R} \sim m_{H_1}/2$ is more efficient as compared to dips at $ M_W/2$ and $M_Z/2$. This is because couplings corresponding to Higgs channels are much stronger (being proportional to $\lambda_i \, v_1$) than the gauge couplings of the W and Z channel annihilations. 

Furthermore, as shown in Fig.~\ref{fig:relic}, the correct relic density points (depicted in magenta) are confined to the parameter space where
$ m_{\eta_2^R} \lsim 100$ GeV. Beyond this region (i.e. $ m_{\eta_2^R}> 100$ GeV), there are no data points corresponding to correct relic abundance. In the latter region, the relic density is under abundant. 
This occurs because once the  $A_4$ symmetry is broken by $\langle\eta_1\rangle= v_2/\sqrt{2}$, the mass parameter $\mu^2_{\eta}$ in Eq.~\eqref{eq:scalarpot} is no longer free parameter, instead, it depends on quartic couplings  $\lambda$s, see Eq.~\eqref{eq:mass_eq}. Thus, achieving a higher mass value for $\eta_2^R$ is possible only by increasing the $\lambda$s. 
However, the larger the value of $\lambda$s (and hence $m_{\eta_2^R} $), the larger the cross-section of DM annihilation to SM particles.  
Consequently, the value of relic density will decrease, and beyond  $m_{\eta_2^R}  \lsim 100$ GeV, DM is always under abundant.
Finally, apart from satisfying the relic abundance, a good candidate for DM should also satisfy various experimental constraints including the collider and direct detection constraints which we discuss in the next section.

\subsubsection*{\textbf{ Collider and Direct detection constraints}}\label{sec:dd}

This section discusses the collider and direct detection (DD) experimental constraints for the DM candidate $\eta_2^R$. The collider constraints namely, LEP (LEP-I, LEP-II) and LHC have a significant impact on the viable parameter space of the DM mass. The LEP constraints impose lower bounds on the masses of dark sector scalar particles, with the most stringent ones being:
\textit{
\begin{itemize}
\item The LEP-I measurements rule out SM-gauge boson decays into dark sector particles~\cite{Cao:2007rm,Gustafsson:2007pc}. This condition leads to the following lower limit on the dark sector scalar masses:
\begin{align} \label{eq:LEP-I}
m_{\eta_{2/3}^{R/I}} + m_{\eta_{2/3}^{\pm}}> M_W, \quad m_{\eta_{2/3}^{R}} + m_{\eta_{2/3}^{I}}> M_Z, \quad m_{\eta_{2/3}^{\pm}} + m_{\eta_{2/3}^{\mp}}> M_Z \;.
\end{align}
\item Chargino searches in LEP-II in context of  singly-charged scalar production $e^+ e^- \rightarrow h^+ h^-$ can be adapted to our analysis to set the limits~\cite{Pierce:2007ut} given by
\begin{align} \label{eq:LEP-II}
m_{\eta_{2/3}^{\pm}}> 70 \hspace{0.10cm} \text{\rm GeV} \; .
\end{align}
\end{itemize}}

The most important LHC limit comes from the possibility of Higgs ``invisible decay" to dark sector particles. The Higgs invisible decay comes through the channel $H_1 \to \eta_{2/3}^{R/I} \eta_{2/3}^{R/I}$. The current bound for the branching ratio of Higgs invisible from LHC data has limit~\cite{ATLAS:2022yvh}: 

\begin{align} \label{eq:LHC}
    BR(H_1 \to inv)<0.145.
\end{align}

Apart from collider constraints, the DM parameter space is also constrained by limits from DD experiments. 
\begin{figure}[h!]
\centering
\includegraphics[width=0.8\textwidth]{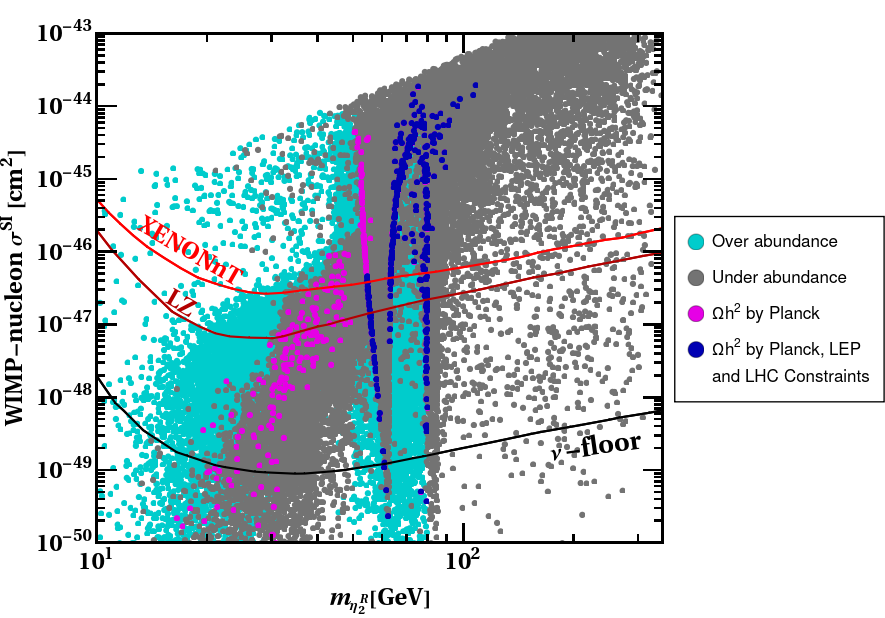}
\caption{\footnotesize \centering Spin-independent WIMP-nucleon cross section for the scalar DM.} 
\label{fig:directdetection}
\end{figure}    
In our model the tree-level spin-independent $\eta_2^R$-nucleon cross section is mediated by the Higgs ($H_1$) and the Z-portals as shown in Fig.~\ref{fig:ddscalar} in Appendix~\ref{feyndiag1}. 
Note that, like the canonical scotogenic model, here also the Higgs portal dominates and the contribution through the Z-portal can be neglected when compared to the Higgs portal. 
We have shown our result for direct detection of the scalar DM candidate $\eta^R_2$ in Fig.~\ref{fig:directdetection}, varying the input parameters as given in Table~\ref{tab:scan1}. 
For consistency, we utilize the same color code used in Fig.~\ref{fig:relic} with cyan, magenta, and gray points representing over, correct, and under abundant relic density, respectively.
The blue color points in Fig.~\ref{fig:directdetection} represent the correct relic density points which also satisfies the collider constraints mentioned in Eqs.~\eqref{eq:LEP-I}, \eqref{eq:LEP-II}, \eqref{eq:LHC} by LEP (LEP-I, LEP-II), and LHC. 

The red and dark red line in Fig.~\ref{fig:directdetection} denotes the latest DD upper bounds from the XENONnT collaboration~\cite{XENON:2023cxc} and LZ collaboration~\cite{LZ:2022lsv}, respectively. 
The black line represents the lower limit corresponding to the ``neutrino floor" from coherent elastic neutrino scattering~\cite{Billard:2013qya}. 
From Fig.~\ref{fig:directdetection}, one can infer that the blue points that lie below the LZ limit are allowed points satisfying the relic density, LEP, LHC and DD constraints. This allowed region is from $55$ GeV to $80$ GeV.  Thus, our model is compatible with all the constraints within the DM mass range of $55\, \text{GeV} \leq m_{\eta^R_2} \leq 80\, \text{GeV}$. 
%

\section{Conclusions}\label{sec:Conclusion}

In this work, we have proposed a novel framework where tiny neutrino masses,  the
two observed mass scales in neutrino oscillation experiments, the leptonic flavor structure, and dark matter can be explained using a single flavor symmetry.  We have demonstrated the framework by explicitly constructing a simple model using the \A4 flavor symmetry. 
The model extends the SM by introducing scalars and fermions that transform as triplets under the $A_4$ symmetry. 
We show that the spontaneous breaking of \A4 symmetry to its \Z2 subgroup leads to a scoto-seesaw hybrid mass mechanism in the neutrino sector, where an interplay between the tree-level type-I seesaw and the one-loop level scotogenic mechanism generates neutrino masses.
Furthermore, the residual \Z2 acts as a scotogenic dark symmetry that stabilizes the DM. 

After conducting numerical analysis, we find that the model aligns with the normal neutrino mass ordering and predicts a small range along with a lower bound on the lightest neutrino mass. The predicted flavor mixing patterns exhibit a generalized $ \mu-\tau $ reflection symmetric flavor structure, consistent with the latest neutrino oscillation data, with  $\delta_{\rm{CP}}$ phase close to $1.5 \pi$ value.
We also discuss the predictions of our model for the neutrinoless double beta decay and beta decay experiments. The results are summarized in Figs.~\ref{fig:NuSector1} - \ref{fig:NuSector2}.

For the dark sector, we show that the experimental constraints coming from neutrino oscillation data rule out the possibility of having fermionic DM, a reflection of the flavor origin of the dark sector. 
For the scalar DM case, the higher mass region is not allowed due to the perturbativity limit of scalar couplings. The observed DM relic abundance can be only explained for the lower mass region $m_{\eta^R_2} \leq 100 $ GeV (see Fig.~\ref{fig:relic}). Further, the imposition of LEP, LHC, and direct detection constraints makes the allowed parameter space for DM mass to be in the $55 \hspace{0.10cm} \text{GeV} \leq m_{\eta^R_2} \leq 80 \hspace{0.10cm} \text{GeV}$ range. 
In summary, our explicit model is highly predictive with its predictions testable in various neutrino sectors as well as dark matter experiments. 

Before ending, we want to emphasize that the model presented in this work serves as a specific example of a broader framework, where an appropriate flavor symmetry with proper residual subgroup can be used to achieve similar results. Thus, various extensions and generalizations of this model are possible. For example, even with \A4 symmetry, this idea can be generalized using a different set of BSM particles and/or assigning the particles the \A4 representations in a different way or by using different one or higher loop neutrino mass models. One can also use other flavor symmetries whose subgroups potentially can stabilize the DM candidate. We plan to discuss some of these possibilities in future works.

\section{Acknowledgements}
\noindent Authors would like to acknowledge the MPT~\cite{Antusch:2005gp}, SARAH~\cite{Staub:2015kfa}, SPheno~\cite{Porod:2011nf} and micrOMEGAS~\cite{Belanger:2014vza} packages which have been used to perform the numerical analysis.
RK acknowledges the funding support by the CSIR SRF-NET fellowship. NN  is supported by the Spanish grants PID2020-113775GB-I00 (AEI/10.13039/501100011033) and Prometeo CIPROM/2021/054 (Generalitat Valenciana). NN also acknowledges the Istituto Nazionale di Fisica Nucleare (INFN) through the ``Theoretical Astroparticle Physics" (TAsP) project. The work of RS is supported by SERB,
Government of India grant SRG/2020/002303. 

RS would like to dedicate this work to his newborn daughter Akanksha whose gestation took less than half the gestation time of this work.
\appendix  
\section{$A_4$ symmetry and scalar potential} \label{app:A4}
\subsection{Transformation rules for  \A4  symmetry and its \Z2 subgroup}
\label{sec:app-A4}
The $A_4$ symmetry is a non-abelian discrete flavor group. It is an even permutation group of four objects. It is also the symmetry group of a regular tetrahedron. It has 12 elements and can be generated by two generators $S$ and $T$ obeying the relations:
\begin{equation} \label{eq:a4genrel}
S^2=T^3=(ST)^3=I.
\end{equation}
In the basis where $S$ and $T$ are real matrices, the generators are given by,
\begin{equation} \label{eq:a4genmat}
S=\left(
\begin{array}{ccc}
1&0&0\\
0&-1&0\\
0&0&-1\\
\end{array}
\right)\,, \quad
T=\left(
\begin{array}{ccc}
0&1&0\\
0&0&1\\
1&0&0\\
\end{array}
\right)\;.
\end{equation}

\A4 has four irreducible representations, three of them are one-dimensional (i.e. three singlets) $1$, $1'$, and $1''$, and one of them is three-dimensional (triplet) $3$.
The multiplication rule for these representations are:
\begin{eqnarray} \label{eq:a4mrule}
&1\times1&=1=1' \times 1'', \quad 1'\times 1'=1'', \quad 1''\times 1''=1',  \nonumber \\
&1 \times 3&=3, \quad 3\times 3= 1+ 1' + 1'' + 3_1 + 3_2 \quad .
\end{eqnarray}
where $3_1$ and $3_2$ denote the two different ways the components of the triplets can be multiplied.
If $a=\left(a_1,a_2,a_3\right)$ and $b=\left(b_1,b_2,b_3\right)$ are two triplets then their  multiplication rules are given as follows~\cite{Ishimori:2010au}:
\begin{equation}\label{eq:pr}
\begin{array}{lll}
\left(ab\right)_1&=&a_1b_1+a_2b_2+a_3b_3\, ,\\
\left(ab\right)_{1'}&=&a_1b_1+\omega a_2b_2+\omega^2a_3b_3\, ,\\
\left(ab\right)_{1''}&=&a_1b_1+\omega^2 a_2b_2+\omega a_3b_3\, ,\\
\left(ab\right)_{3_1}&=&\left(a_2b_3,a_3b_1,a_1b_2\right)\, ,\\
\left(ab\right)_{3_2}&=&\left(a_3b_2,a_1b_3,a_2b_1\right)\,,
\end{array}
\end{equation}
where $\omega$ is the cubic root of unity.  
%

The $A_4$ group has a \Z2 subgroup as mentioned before. If the \A4 group is broken spontaneously by VEV of a triplet scalar $\varphi$ with  VEV alignment $\langle \varphi \rangle \sim \frac{1}{\sqrt{2}}(v,0,0)$, then the \Z2 group remains unbroken~\cite{Boucenna:2011tj}. This happens because the generator $S$ remains invariant i.e.
\begin{equation} \label{eq:a4genvev}
S \langle \varphi \rangle \equiv \left(
\begin{array}{ccc}
1&0&0\\
0&-1&0\\
0&0&-1\\
\end{array}
\right) \left(
\begin{array}{ccc}
\frac{v}{\sqrt{2}}\\
0\\
0 \\
\end{array}
\right)
=
\left(
\begin{array}{ccc}
\frac{v}{\sqrt{2}}\\
0\\
0 \\
\end{array}
\right)
\equiv \langle \varphi \rangle .
\end{equation}
leaving the \Z2 subgroup unbroken.

In such a case any other field transforming as one of the singlets of $A_4$ will be even $(+)$ under the residual \Z2 subgroup. For a triplet of \A4, $\Psi \equiv \left(a_1,a_2,a_3 \right)^T$, under $S$ transformation we have:
\begin{align} \label{eq:a4z2con}
S \Psi =\left(
\begin{array}{ccc}
1&0&0\\
0&-1&0\\
0&0&-1\\
\end{array}
\right) \left(
\begin{array}{ccc}
a_1\\
a_2\\
a_3 \\
\end{array}
\right)= \left(
\begin{array}{ccc}
a_1\\
-a_2\\
-a_3 \\
\end{array}
\right)
\end{align}
This means that after $A_4 \to$ \Z2, the first component of the triplet field $\Psi$ transforms as even $(+)$ under \Z2 while the other two components transform as odd $(-)$ under \Z2. This fact is crucial to our work.

In our model, we have two triplets $N$ and $\eta$. Once \A4 symmetry is broken, the transformation of their components under the \Z2 residual symmetry is given by:
\begin{equation} \label{eq:a4fielddec}
\begin{array}{lcrlcr}
N_1 &\to& +N_1\,,\quad& \eta_1 &\to& +\eta_1 \, ,\\  
N_{2,3} &\to& -N_{2,3}\,,\quad& \eta_{2,3} &\to& -\eta_{2,3} \;.
\end{array}
\end{equation}
The remaining fields are all \Z2 even, because they are singlets of \A4. Thus after \A4 $\to$ \Z2 symmetry breaking, the particles running inside the loop all have \Z2 odd charges while all SM particles are \Z2 even. This allows us to use the unbroken \Z2 subgroup as the dark symmetry of the scotogenic model. 

\subsection{Scalar Potential}
\label{sec:app-scalar}
Now moving to the scalar potential, we have two scalars $\Phi$ and $\eta$, singlet and triplet of $A_4$ respectively. The $SU(3)_C\otimes SU(2)_L\otimes U(1)_Y \otimes A_4$ invariant scalar potential can be written as:
\begin{align}
\label{eq:pot1}
V&=\mu_{\Phi}^2 \Phi^\dagger \Phi 
+ \mu_\eta^2[\eta^\dagger\eta]_1 + \lambda_1 (\Phi^\dagger \Phi)^2+\lambda_2 [\eta^\dagger\eta]_1^2
+\lambda_3 [\eta^\dagger\eta]_{1^{\prime}}[\eta^\dagger\eta]_{1^{\prime\prime}} +\lambda_4 [\eta^\dagger\eta^\dagger]_{1^\prime}[\eta\eta]_{1^{\prime\prime}}
\nonumber \\
&+\lambda_{4^\prime}[\eta^\dagger\eta^\dagger]_{1^{\prime\prime}}[\eta\eta]_{1^\prime}
+\lambda_5[\eta^\dagger\eta^\dagger]_{1}[\eta\eta]_{1}
+\lambda_6\left([\eta^\dagger \eta]_{3_{1}}[\eta^\dagger \eta]_{3_{1}}+h.c.\right)+\lambda_7 [\eta^\dagger \eta]_{3_{1}}[\eta^\dagger \eta]_{3_{2}} 
\nonumber \\
&+\lambda_8 [\eta^\dagger \eta^\dagger]_{3_{1}}[\eta \eta]_{3_{2}} 
+\lambda_9 [\eta^\dagger \eta]_1 (\Phi^\dagger \Phi) 
+\lambda_{10}[\eta^\dagger \Phi]_3[\Phi^\dagger \eta]_3 +\lambda_{11}\left([\eta^\dagger\eta^\dagger]_{1}\Phi \Phi+h.c.\right) 
\nonumber \\
&+\lambda_{12}\left([\eta^\dagger\eta^\dagger]_{3_{1}}[\eta \Phi]_3+h.c.\right) 
+\lambda_{13}\left([\eta^\dagger\eta^\dagger]_{3_{2}}[\eta \Phi]_3+h.c.\right) 
+\lambda_{14}\left([\eta^\dagger \eta]_{3_{1}}[\eta^\dagger \Phi]_3+h.c.\right) \nonumber\\
&+\lambda_{15}\left([\eta^\dagger \eta]_{3_{2}} [\eta^\dagger \Phi]_3+h.c.\right) \;, 
\end{align}
where $[...]_p$;  $p=1,1',1'',3,3_1,3_2$ denote the \A4 transformation of enclosed fields.  

To see how many independent $\lambda$'s appear in the potential, we expand terms in their $A_4$ components.
Upon expanding, it becomes evident that certain terms are common across different couplings, allowing us to combine them as written in Eq.~\eqref{eq:pot3}. This can be understood from the $A_4 \otimes SU(2)_L$ group multiplication. The number of invariants under $A_4 \otimes SU(2)_L$ is determined by the following multiplication:
\begin{align*}
     (\eta \eta) \sim (3,2) \times (3,2)= \left( 1 + 1' + 1^{\prime \prime}+3_1,3 \right ) + \left( 3_2,1\right).
\end{align*}
As a result, we end up with five independent invariants of the type $\left( \eta^\dagger \eta\right) (\eta^\dagger  \eta)$~\cite{Buskin:2021eig}. Also, there are  four invariants of the type $\left( \eta^\dagger \eta \right) (\Phi^\dagger \Phi)$ or $\left( \eta^\dagger \Phi \right)^2 $ and two of $\left( \eta^\dagger \eta \right) (\eta^\dagger \Phi)$ type, as given in Eq.~\eqref{eq:pot3}. 

The potential is CP conserving if all couplings are taken to be real and  $\lambda_{4^\prime}=\lambda_4$.
This can be written in combined form as
\begin{align}
\label{eq:pot3}
V&=\mu_{\Phi}^2 \Phi^\dagger \Phi 
+ \mu_\eta^2\left(\eta_1^\dagger\eta_1 + \eta_2^\dagger\eta_2 + \eta_3^\dagger\eta_3 \right)+\lambda_1 \left(\Phi^\dagger \Phi\right)^2+ \bm{{\Lambda}_2} \left( (\eta_1^\dagger\eta_1)^2 + (\eta_2^\dagger\eta_2)^2 + (\eta_3^\dagger\eta_3)^2 \right)\nonumber \\&
+ \bm{{\Lambda}_3} \left[\left(\eta_1^\dagger\eta_1 \right)\left(\eta_2^\dagger\eta_2 \right)+\left(\eta_2^\dagger\eta_2 \right)\left(\eta_3^\dagger\eta_3 \right)+\left(\eta_3^\dagger\eta_3 \right)\left(\eta_1^\dagger\eta_1 \right) \right] + \bm{{\Lambda}_4}\left[ \left(\eta_1^\dagger\eta_2\right)^2 + \left(\eta_2^\dagger\eta_3 \right)^2 + \left(\eta_3^\dagger\eta_1 \right)^2 + h.c. \right ] 
\nonumber  \\&+\lambda_7 \left[\left(\eta_2^\dagger\eta_3 \right)\left(\eta_3^\dagger\eta_2 \right)+ \left(\eta_3^\dagger\eta_1 \right)\left(\eta_1^\dagger\eta_3 \right)+ \left( \eta_1^\dagger\eta_2 \right)\left(\eta_2^\dagger\eta_1 \right)\right]
+\lambda_9\left(\eta_1^\dagger\eta_1 + \eta_2^\dagger\eta_2 + \eta_3^\dagger\eta_3 \right)\left(\Phi^\dagger \Phi\right)
\nonumber \\
&+\lambda_{10}\left[\left(\eta_1^\dagger \Phi\right)\left(\Phi^\dagger \eta_1\right) + \left(\eta_2^\dagger \Phi \right)\left(\Phi^\dagger \eta_2 \right) + \left(\eta_3^\dagger \Phi \right)\left(\Phi^\dagger \eta_3 \right)\right] +\lambda_{11}\left[\left(\eta_1^\dagger \Phi \right)^2 + \left(\eta_2^\dagger \Phi \right)^2 + \left(\eta_3^\dagger \Phi \right)^2 +h.c.\right] 
\nonumber \\
&+ \bm{{\Lambda}_{12}}\left[\left(\eta_3^\dagger\eta_1 \right)\left(\eta_2^\dagger\Phi \right)+\left(\eta_1^\dagger\eta_2 \right)\left(\eta_3^\dagger\Phi \right)+\left(\eta_2^\dagger\eta_3 \right)\left(\eta_1^\dagger\Phi \right) +h.c.\right] 
\nonumber \\
&+ \bm{{\Lambda}_{13}} \left[\left(\eta_2^\dagger\eta_1 \right)\left(\eta_3^\dagger\Phi \right)+\left(\eta_3^\dagger\eta_2 \right)\left(\eta_1^\dagger\Phi \right)+\left(\eta_1^\dagger\eta_3 \right)\left(\eta_2^\dagger\Phi \right)+h.c.\right]  \;.
\end{align}
where $\bm{{\Lambda}_2} = \left(\lambda_2+ \lambda_3+ 2\lambda_4+ \lambda_5\right) \equiv \Lambda$ (see Eq.\eqref{eq:coupl}),  $\bm{{\Lambda}_3} = \left(2\lambda_2- \lambda_3+ \lambda_8\right), \ \bm{{\Lambda}_4} = \left(\lambda_5+ \lambda_6- \lambda_4\right), \ \bm{{\Lambda}_{12}} = (\lambda_{12}+\lambda_{14})$, and $\bm{{\Lambda}_{13}} = (\lambda_{13}+\lambda_{15})$.

Furthermore, the potential of Eq.~\eqref{eq:pot3} can be simplified in $SU(2)_L$ components and the mass spectrum of scalars can be computed as discussed in Sec.~\ref{subsec:massspect}.
\section{Fermionic dark matter} 
\label{app:fermionicDM} 

As we have already discussed in Sec.~\ref{sec:scalarsec}, in our models all the scalar masses have an upper bound $\lesssim 600$ GeV. For a dark sector particle to be a good DM candidate its mass has to be less than the other dark sector particles. Therefore, even fermionic DM candidates in our model should have masses $ \lesssim 600$ GeV.
Here, we have two potential fermionic DM candidates $N_2$ and  $N_3$ which are odd under \Z2 and are nearly mass degenerate. For the sake of definiteness, we have chosen $N_2$ as the lightest dark sector particle and hence a potential DM candidate with the following conditions:

\begin{align} \label{eq:fermcond}
M_2\leq M_3, \ m_{\eta^R_{2}}, m_{\eta^R_{3}}, m_{\eta^I_{2}}, m_{\eta^I_{3}},m_{\eta^{\pm}_{2}}, m_{\eta^{\pm}_{3}} \ \lesssim 600 \ \rm{GeV}.
\end{align}
Choosing $N_3$ as a DM candidate will imply $M_3 \leq M_2$ but will not change any of the salient points of the analysis.
The numerical analysis has been done following the Table~\ref{tab:scan3} and Eq.~\eqref{eq:fermcond}.
\begin{table}[h!t]
\centering
\begin{tabular}{|c|c|c|c|}
\hline
Parameters & Range &Parameters & Range\\
\hline
$\lambda_1$ & $[10^{-3}, \sqrt{4\pi}]$ & $\lambda_{12,13,14,15}$ & $[10^{-8}, 10^{-2}]$ \\
$|\lambda_{2,3...,10}|$ & $[10^{-6}, \sqrt{4\pi}]$ &$M_{1,2,3}$ (in GeV) & $[10, 600]$ \\
$\lambda_{11}$ & $-[10^{-6}, \sqrt{4\pi}]$ & $y_i$ &$[10^{-8}, 10^{-6}]$\\
\hline
\end{tabular}
\caption{\footnotesize \centering Value range for the numerical parameter scan in the dark sector for fermionic DM.}
 \label{tab:scan3}
\end{table} 
%

\subsubsection*{\textbf{Relic density}} 

\begin{figure}[h!t]
\centering
\includegraphics[width=0.48\textwidth]{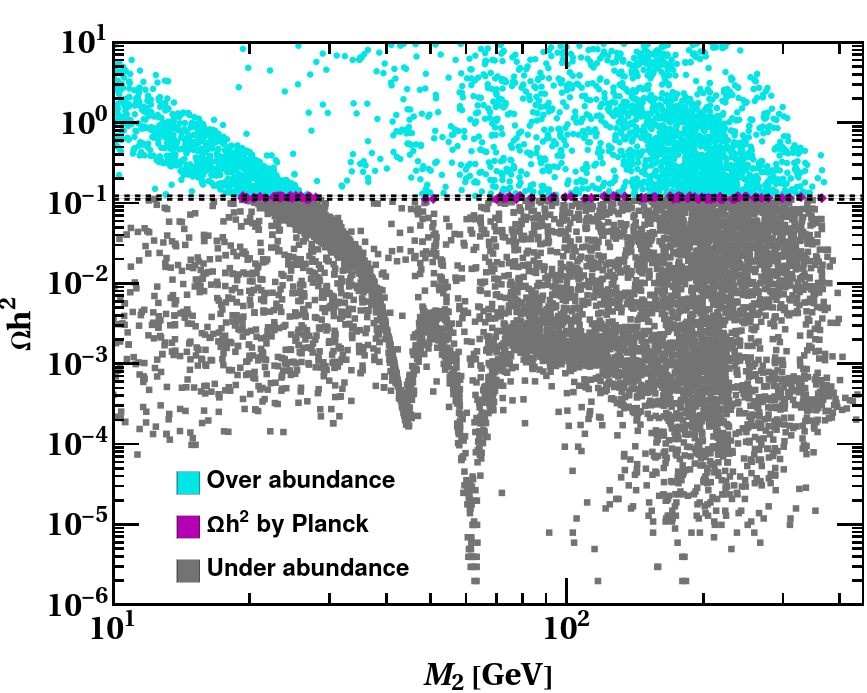}
\includegraphics[width=0.48\textwidth]{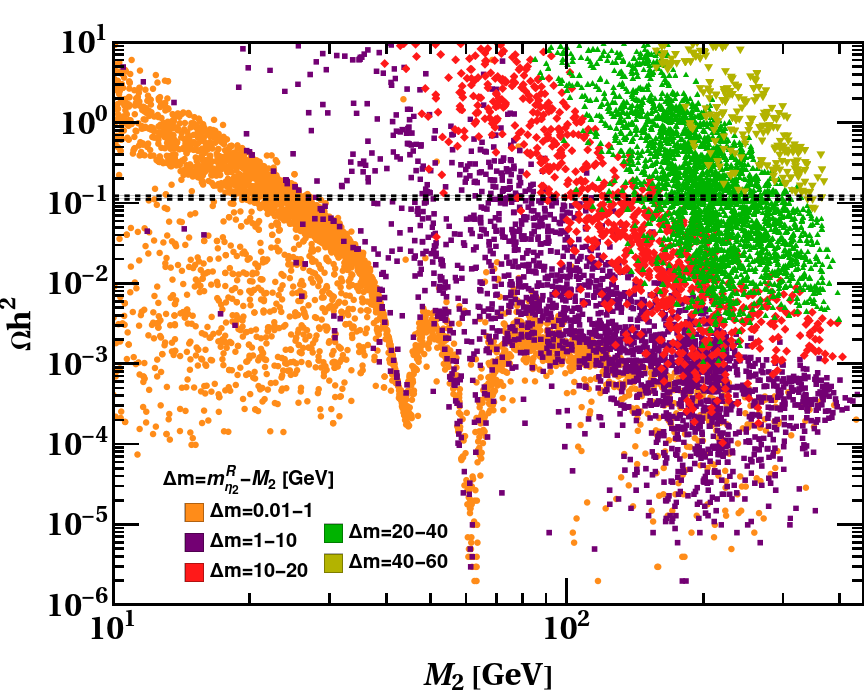}
\caption{\footnotesize Relic density as a function of fermionic DM mass $ M_2 $ has been presented in both panels. In the left panel, cyan, magenta, and gray points depict over, observed, and under abundance of the DM relic density, respectively. 
The narrow band signifies the latest Planck satellite data~\cite{Planck:2018vyg}.
The same has been shown in the right panel, where the behavior of relic density has been shown with respect to the mass difference between the lightest dark sector scalar and fermionic DM (i.e., $ \Delta m = m_{\eta^R_2} - M_2 $) using the orange, purple, red, green, and yellow points, where values of $ \Delta m  $ are marked in the plot. See the main text for detailed discussion.}
\label{fig:FDMSector}
\end{figure}

We show our results for DM relic density ($\Omega h^2$) as a function of the fermionic DM mass $ M_2 $ in Fig.~\ref{fig:FDMSector}. 
The correct relic density can be obtained for the $ N_2 $ mass as heavy as $ \approx 400 $ GeV. In the left panel of Fig.~\ref{fig:FDMSector} we have shown correct relic density points in magenta color along with the over abundant and under abundant points shown in cyan and gray colors respectively. The right panel of Fig.~\ref{fig:FDMSector} illustrates the behavior of relic density based on the mass difference of   $N_2$ and the lightest dark sector scalar $\eta^R_{2}$. One can notice that for large mass difference $m_{\eta^R_{2}}-M_2=\Delta m \sim 60$ GeV most of the points (yellow color points in the right panel of Fig.~\ref{fig:FDMSector}) are over abundant. This pattern of relic density is due to the different annihilation and co-annihilation modes of the $N_2$, see Appendix~\ref{feyndiag2}. 

Annihilation and co-annihilation channels which determine the observed DM relic density are given in Appendix~\ref{feyndiag2} in Figs.~\ref{fig:anni1}, \ref{fig:coanni}, \ref{fig:coanni1} and \ref{fig:coanni2}. When the masses of dark sector scalars $\eta_{2}$ and $\eta_{3}$ significantly exceed that of the DM candidate $N_2$, only annihilation (co-annihilation) channel $N_2N_{2/3} \rightarrow l_i^{\mp}l_j^{\pm},\nu_i\nu_j$ play significant role in the relic density computation. These interactions are mediated by the heavy dark sector scalars, resulting in relatively small cross sections. The consequence of these small cross sections is a high relic density, leading to an over abundance of DM in most cases where there is a significant mass difference. While it is possible to enhance these cross sections a bit by considering large Yukawa couplings ($y_i$),  it is worth noting that Yukawa couplings ($y_i$) are also bounded by their perturbativity limits as well as by the requirement of obtaining small masses for neutrinos. Thus for larger $\Delta m$ correct relic density cannot be obtained for any value of the Yukawa couplings. Changing the mass differences between other dark sector scalars (besides $\eta^R_{2}$) and $N_2$ also does not help and the conclusion will remain the same.  
In scenarios marked by low mass differences between dark sector scalars and DM candidates, additional annihilation and co-annihilation channels become significant along with the $N_2N_{2/3} \rightarrow l_i^{\mp}l_j^{\pm},\nu_i\nu_j$ channels.  When the mass difference, denoted as $\Delta m$, is relatively small, all these annihilation and co-annihilation channels shown in Figs.~\ref{fig:anni1},~\ref{fig:coanni},~\ref{fig:coanni1},~\ref{fig:coanni2} contribute significantly, leading to a higher cross section compared to scenarios with large $\Delta m$. In the latter case, the primary contribution arises solely from the  $N_2N_{2/3} \rightarrow l_i^{\mp}l_j^{\pm},\nu_i\nu_j$ channel. This enhancement in the cross section for small mass differences results in lower relic density values as well as correct relic density as shown in the right panel of Fig.~\ref{fig:FDMSector}. For mass differences $\Delta m \leq 40$ GeV, a considerable number of data points fall inside the  $3 \sigma$ range for the  DM relic density reported by the Planck satellite data~\cite{Planck:2018vyg}.

\subsubsection*{ \textbf{Direct detection}}
Here, we will briefly discuss the direct detection process for fermionic DM case in our model. Notably, there exists no direct coupling between the fermionic DM $N_2$ and quarks. Hence, at the tree-level, no interaction occurs between $N_2$ and quarks. Nonetheless, at the one-loop level, $N_2$ can couple with quarks through Z boson, photon ($\gamma$), and Higgs boson ($H_1$) as depicted in Fig.~\ref{fig:DD} in Appendix~\ref{feyndiag2} leading to a non-zero DM-nucleon cross section. 

The spin-dependent cross section per nucleon is given by the following expression~\cite{Ibarra:2016dlb}. 
\begin{align*}
\sigma_{SD}\sim 10^{-4} \rm{pb} \left(\frac{\mathit{y_i}}{3}\right)^4 \mathcal{G}_2\left(\frac{M^2_2}{m^2_{\eta}}\right)^2
\end{align*}
where, $\mathcal{G}_2(x)$ is loop function given in Ref.~\cite{Ibarra:2016dlb} and  for our model it has value $\mathcal{G}_2\left(\frac{M^2_2}{m^2_{\eta}}\right)^2\sim1$.

The neutrino oscillation constraints imply\footnote{As discussed in the next section, the fermionic DM cannot simultaneously satisfy $\Delta m^2_{\rm{atm}}$ and  $\Delta m^2_{\rm{sol}}$ constraints. We are taking Yukawas corresponding to the green points in the right panel of Fig.~\ref{masssquare} which comes closest to simultaneously satisfying both mass square differences.} that, for this case, all Yukawas in our model are of the order, $y_i\sim 10^{-8} - 10^{-6}$, see next section.
Therefore,
\begin{align} \label{eq:ddcrsec}
\sigma_{SD}\sim 10^{-38} - 10^{-30} \, \rm{pb}.
\end{align} 
Also, the spin-independent cross section per nucleon, $\sigma_{SI} \propto y_i^4$ as discussed in~\cite{Ibarra:2016dlb}. Hence, $\sigma_{SI}$ will also be equally small.

Given that direct detection in our model is only possible at the one-loop level and taking into account the small Yukawa couplings ($y_i \sim 10^{-8} - 10^{-6}$), the resulting spin-dependent (independent) cross section $\sigma_{SD}$ ($\sigma_{SI}$) is exceedingly tiny. Consequently, this tiny cross section trivially satisfies the constraints from the direct detection experiments~\cite{XENON:2023cxc,LZ:2022lsv}.
%
\subsubsection*{\textbf{Neutrino oscillations constraints for fermionic DM}}

\begin{figure}[h!]
\centering
\includegraphics[width=0.48\textwidth]{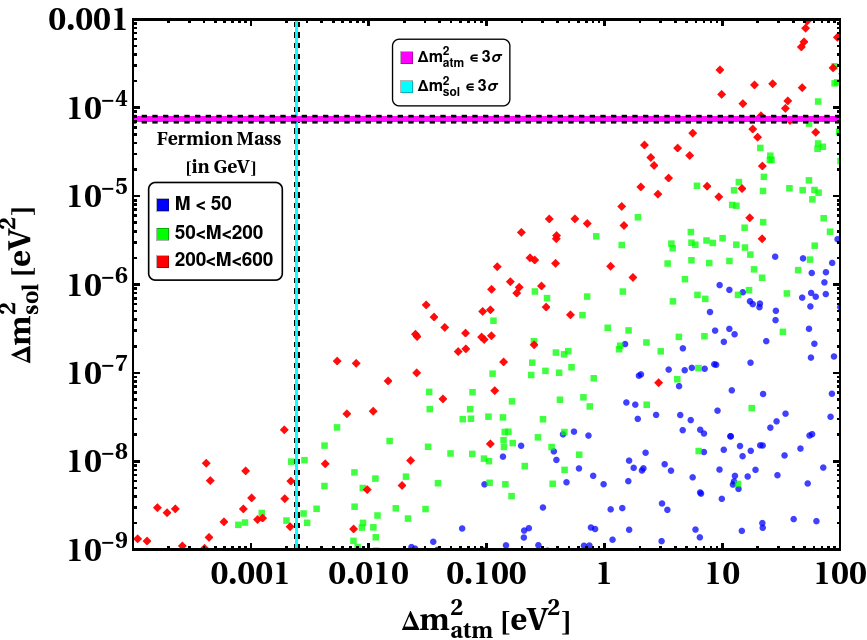}
\includegraphics[width=0.47\textwidth]{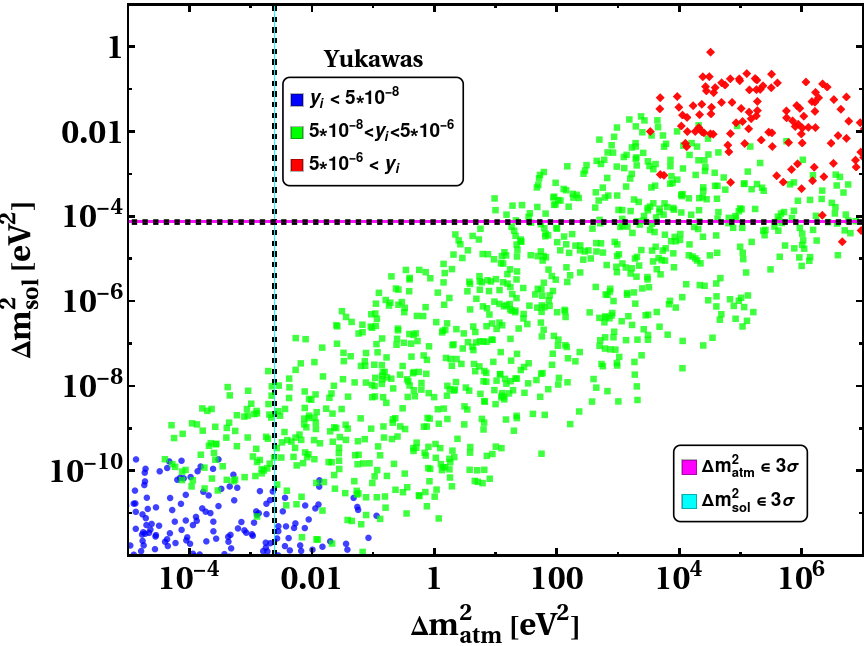}
\caption{\footnotesize Correlation between mass-squared differences ($\Delta m^2_{\rm{atm}}$ and  $\Delta m^2_{\rm{sol}}$) of neutrino oscillation in fermionic DM case. \textbf{Left:} Correlation dependency on fermionic DM mass. \textbf{Right:} Correlation dependency on Yukawa couplings. }
\label{masssquare}
\end{figure}

Although fermionic DM satisfies all DM constraints, this scenario is not compatible with the neutrino oscillation data. As discussed previously, the dark \Z2 symmetry in our model emerges from the breaking of the \A4 symmetry. Thus the neutrino and dark sectors are intimately tied together with the same dark sector particles and Yukawa couplings playing important roles in both sectors. Furthermore, the scalar masses have an upper bound ($\sim 600$ GeV),  hence for $N_2$ to be fermionic DM, its mass has to be within this limit. 
In Fig.~\ref{masssquare}, we have shown that how  $\Delta m^2_{\rm{atm}}$ and  $\Delta m^2_{\rm{sol}}$ are related with each other. All points in Fig.~\ref{masssquare} lie within the $3 \sigma$ allowed range of mixing angles from global-fit data~\cite{deSalas:2020pgw}. The band shown in cyan and magenta color depicts the $3 \sigma$ allowed values of $\Delta m^2_{\rm{atm}}$ and  $\Delta m^2_{\rm{sol}}$ respectively~\cite{deSalas:2020pgw}. 

The left panel of Fig.~\ref{masssquare} depicts the behavior of two mass-squared differences with the mass of fermionic DM. 
We notice that for lower values of fermionic DM mass, we can not satisfy any mass constraint while for higher values of fermionic DM mass, we can satisfy only one mass-squared constraint at a time. In the right panel of Fig.~\ref{masssquare} we have shown the behavior of mass-squared differences in terms of Yukawa couplings. Hence, from Fig.~\ref{masssquare} one can infer that within the allowed mass region of fermion $N_2$ (up to $\sim 600$ GeV), we are not able to simultaneously satisfy the two mass-squared differences $\Delta m^2_{\rm{atm}}$ and  $\Delta m^2_{\rm{sol}}$ of neutrino oscillation. The results remain the same if $N_3$ is taken as the fermionic DM.
In fact for any perturbative value of Yukawa couplings and masses of dark fermions  $\leq 10^5$ GeV, the two oscillation length scales $\Delta m^2_{\rm{atm}}$ and  $\Delta m^2_{\rm{sol}}$ cannot be simultaneously obtained within their respective experimental 3$\sigma$ ranges.  Thus, in our model, the neutrino oscillation constraints can only be satisfied for scalar DM.

This feature inherent to our model serves to distinguish it from conventional scotogenic as well as the scoto-seesaw models, where both scalar and fermionic DM are  equally viable. Within the context of our model, the breaking of the $A_4$ symmetry plays a crucial role in having an upper bound on the scalar masses. Hence, for a stable fermionic DM, its mass has to be less than dark sector scalars. Within this limit of fermionic DM mass, the two mass scales of neutrino oscillation data can not be satisfied simultaneously. Therefore, the possibility of fermionic DM in our model is ruled out taking neutrino oscillations data into account. This outcome emphasizes the significance of  \A4 flavor symmetry within our model, distinguishing it from the scotogenic and scoto-seesaw model and their variants. The utilization of this flavor symmetric approach yields unique implications for the DM candidate viability in our model, allowing only for the scalar DM.
\section{Other possible representation of lepton doublets $L_i$ under $A_4$}\label{app:OtherRe}
In this section, we discuss the other possible representations of $A_4$ for the lepton doublets $L_i$. In our current model we have assigned  $L_i$ to $(1, 1', 1'')$ charges under $A_4$. The other possible scenarios for $L_i$ are as follows
\begin{align}
    L_i \sim (1, 1'', 1'),\ (1', 1, 1''),\ (1'', 1, 1'),\ (1', 1'', 1),\ (1'', 1', 1)
\end{align}
To keep the charged lepton mass matrix diagonal,  $e_{R_i}$ should have the same representation as $L_i$. Since the singlets $1'$ and $1''$ of $A_4$ are conjugates of each other, only two distinct configurations, other than the one presented in the main text, are $(1', 1, 1''), \ (1', 1'', 1)$. The final form of the light neutrino mass matrix for these two possibilities are:
\begin{itemize}
    \item \textbf{II. For $L_i \sim (1', 1, 1'')$} case: The final neutrino mass matrix from the tree and loop contribution has the following structure 
    \begin{align}\label{eq:numatii}
        m^{(II)}_{\nu} =
\begin{pmatrix}
y^2_1 \left(d_3 - z\right) & y_1y_2\left(d_2 - z\right) & y_1y_3\left(d_1 - z\right)  \\
\ast & y^2_2 \left(d_1 - z\right) & y_2y_3 \left(d_3 - z\right) \\
\ast & \ast & y^2_3 \left(d_2 - z\right) \\
\end{pmatrix} \;.
\end{align}
 \item \textbf{III. For $L_i \sim (1', 1'', 1)$} case: For this representation, the final neutrino mass matrix has the following structure 
    \begin{align} \label{eq:numatiii}
        m^{(III)}_{\nu} =
\begin{pmatrix}
y^2_1 \left(d_3 - z\right) & y_1y_2\left(d_1 - z\right) & y_1y_3\left(d_2 - z\right)  \\
\ast & y^2_2 \left(d_2 - z\right) & y_2y_3 \left(d_3 - z\right) \\
\ast & \ast & y^2_3 \left(d_1 - z\right) \\
\end{pmatrix} \;.
\end{align}
\end{itemize}
where $z= \frac{v_2^2}{2M}$ and $d_i$, $(i=1,2,3)$ are as defined in Eq.~\eqref{eq:dvalues}. 

Here, we would like to mention that the structure of the matrices in Eqs.~\eqref{eq:numatii} and \eqref{eq:numatiii} differ significantly from the matrix in Eq.~\eqref{eq:NuMassMatrix}. A key distinction arises in the $y_2=y_3$ limit, where matrix~\eqref{eq:NuMassMatrix} resembles the exact $\mu-\tau$ reflection symmetry~\cite{Harrison:2002et}. However, the alternative matrices in~\eqref{eq:numatii} and \eqref{eq:numatiii} fail to exhibit this structure under any particular limit. This highlights the importance of representation choice, contrary to the assertion in Ref.~\cite{Bonilla:2023pna}, where the results obtained by the authors are incorrect.
We have analyzed the neutrino mass matrices provided in ~\eqref{eq:numatii} and \eqref{eq:numatiii} numerically as well and found that indeed numerically as well their predictions differ from $L_i \sim (1, 1', 1'')$. 
\section{Feynman diagrams for  relic density and direct detection}\label{app:DM-DD}
In this section, we show the possible Feynman diagrams that play a role in the determination of the relic density of DM for both scenarios, scalar DM $\eta_2^R$ and fermionic DM $N_2$. In addition, we have included Feynman diagrams for direct detection of DM in both of these cases. The inclusion of these Feynman diagrams serves to assist our understanding of the dark sector results presented in Sec.~\ref{sec:DarkSector} and in Appendix~\ref{app:fermionicDM}. 
\subsection{ Feynman diagrams for scalar DM} \label{feyndiag1}
\begin{figure}[!h] 
\centering
\includegraphics[width=0.82\textwidth]{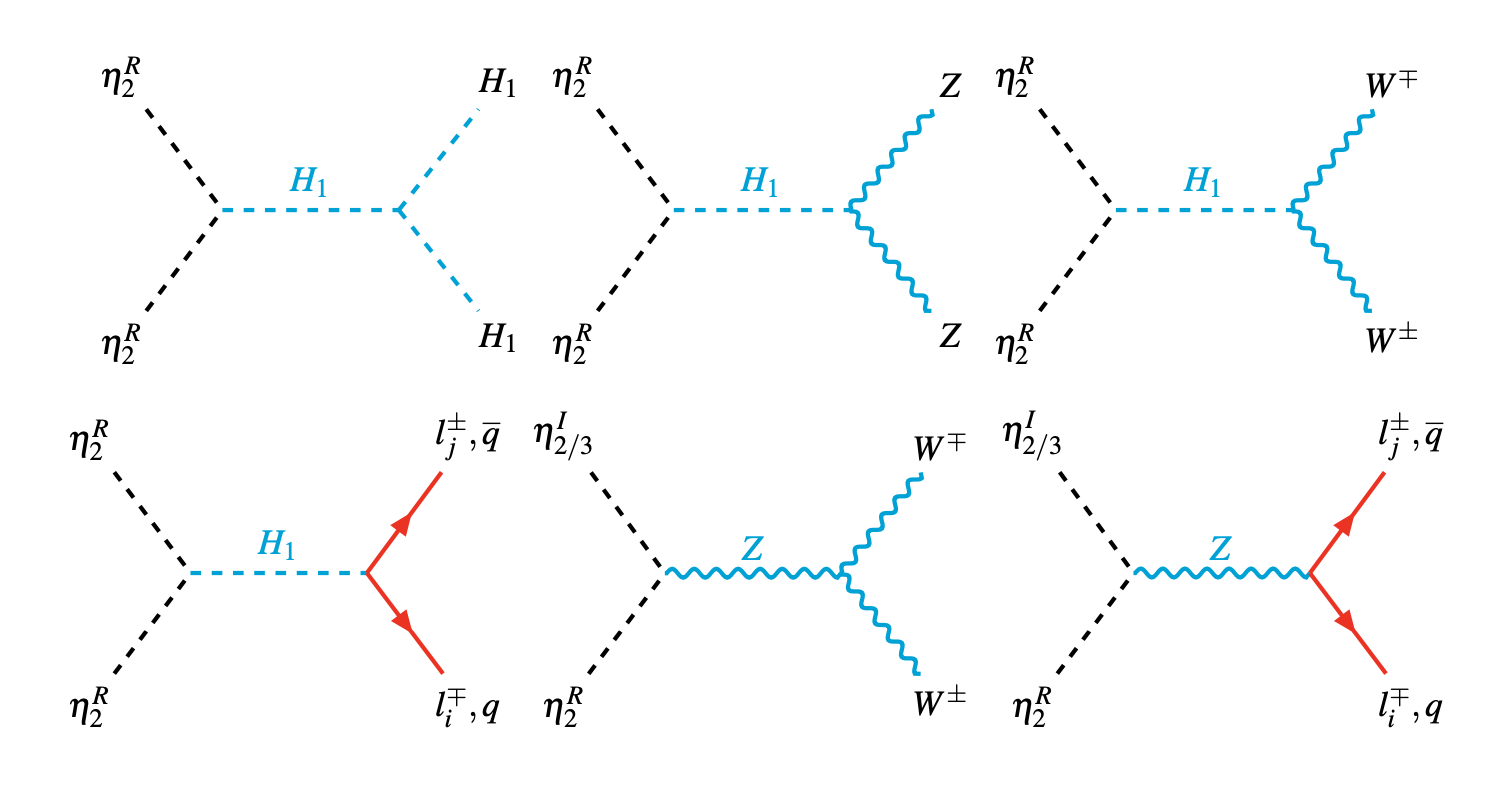}
\caption{\footnotesize \centering Annihilation and co-annihilation channel for scalar DM $\eta^R_2$.}
\label{fig:scanni1}
\end{figure}

\begin{figure}[h!] 
\centering
\includegraphics[width=0.75\textwidth]{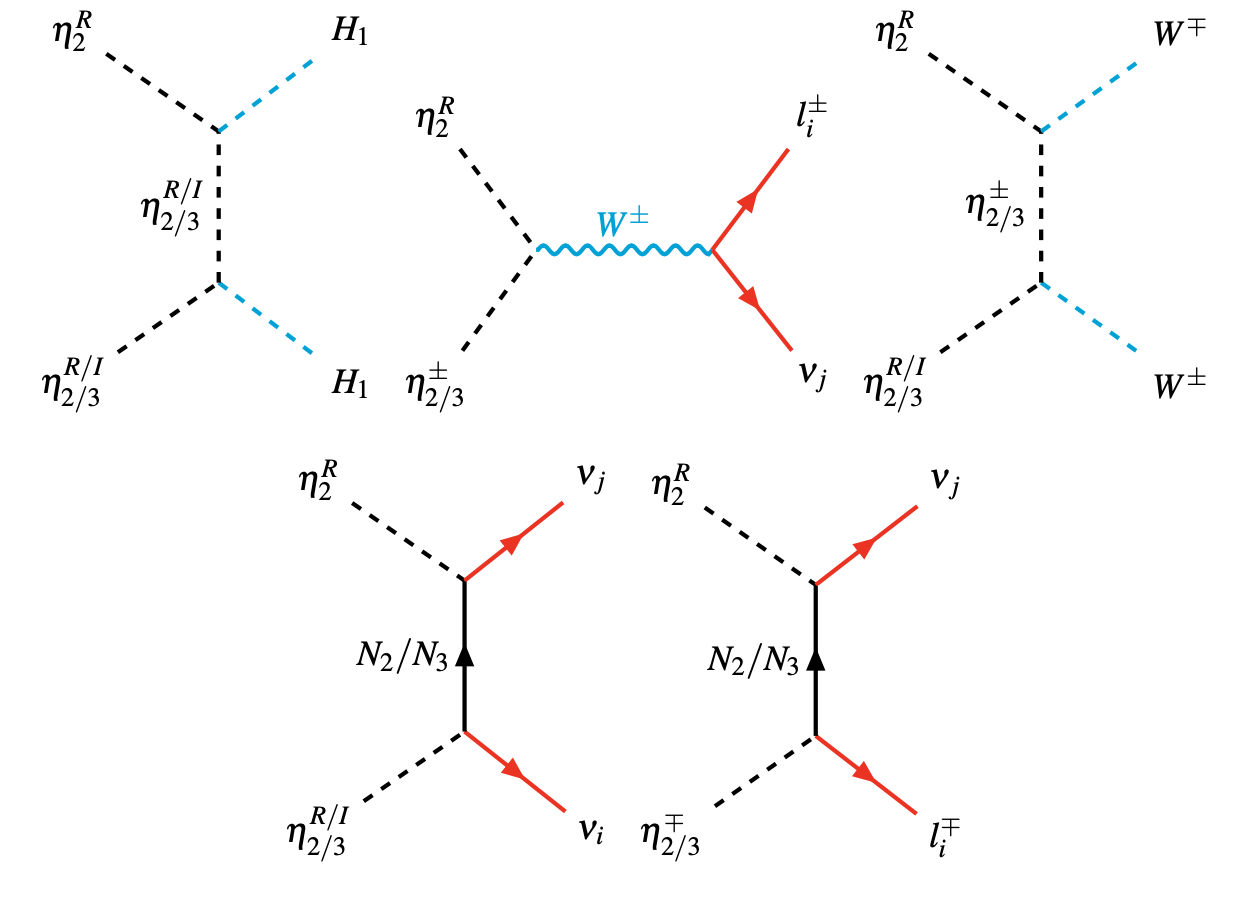}
\caption{\footnotesize \centering Annihilation and co-annihilation channel for scalar DM $\eta^R_2$.}
\label{fig:scanni2}
\end{figure}

\begin{figure}[h!]
\centering
\includegraphics[width=0.45\textwidth]{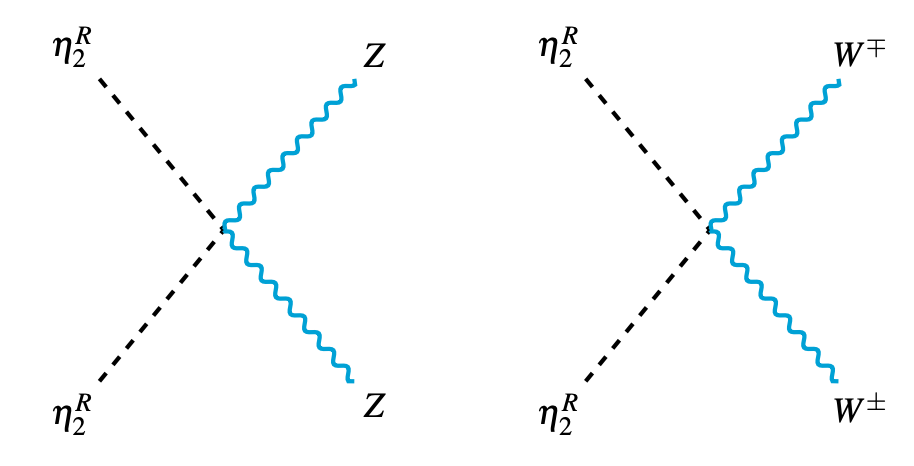}
\caption{\footnotesize \centering Annihilation via quartic couplings for scalar DM $\eta^R_2$.}
\label{fig:quartic}
\end{figure}

\begin{figure}[h!]
\centering
\includegraphics[width=0.45\textwidth]{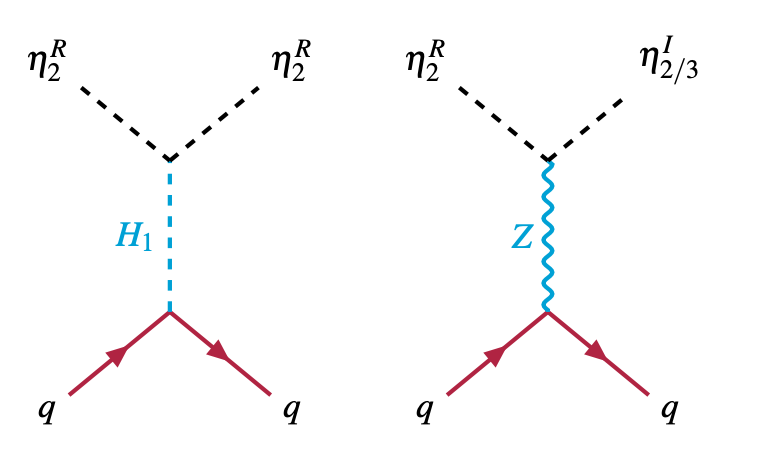}
\caption{\footnotesize \centering Feynman diagrams  of scalar DM $\eta^R_2$
for the direct detection.}
\label{fig:ddscalar}
\end{figure}

In Figs.~\ref{fig:scanni1}, \ref{fig:scanni2} and \ref{fig:quartic}, we have shown possible Feynman diagrams corresponding to the annihilation and co-annihilation channels responsible for determining the relic density of scalar DM $\eta^R_2$. Fig.~\ref{fig:scanni1} illustrates the s-channel processes of DM annihilation and co-annihilation, mediated by the Higgs boson $H_1$, W and Z bosons, along with other potential channels. On the other hand, Fig.~\ref{fig:quartic} focuses on the annihilation of DM into gauge bosons through quartic couplings. In Fig.~\ref{fig:ddscalar}, we have shown the direct detection prospects of scalar DM $\eta^R_2$ mediated by Higgs boson $H_1$ and Z boson.
\subsection{ Feynman diagrams for fermionic DM } \label{feyndiag2} 
\begin{figure}[H]
\centering
\includegraphics[width=0.85\textwidth]{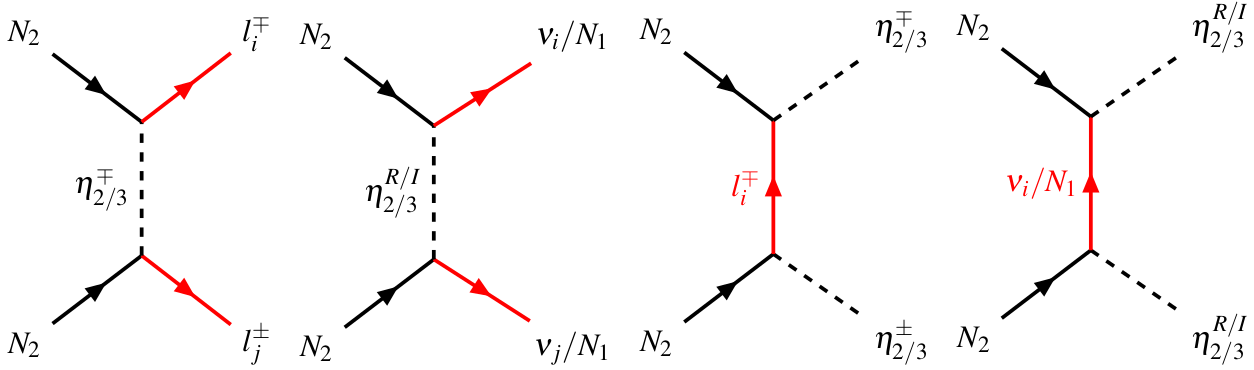}
\caption{\footnotesize \centering Annihilation channels for fermionic DM.}  
\label{fig:anni1}
\end{figure}

\begin{figure}[h!]
\centering
\includegraphics[width=0.9\textwidth]{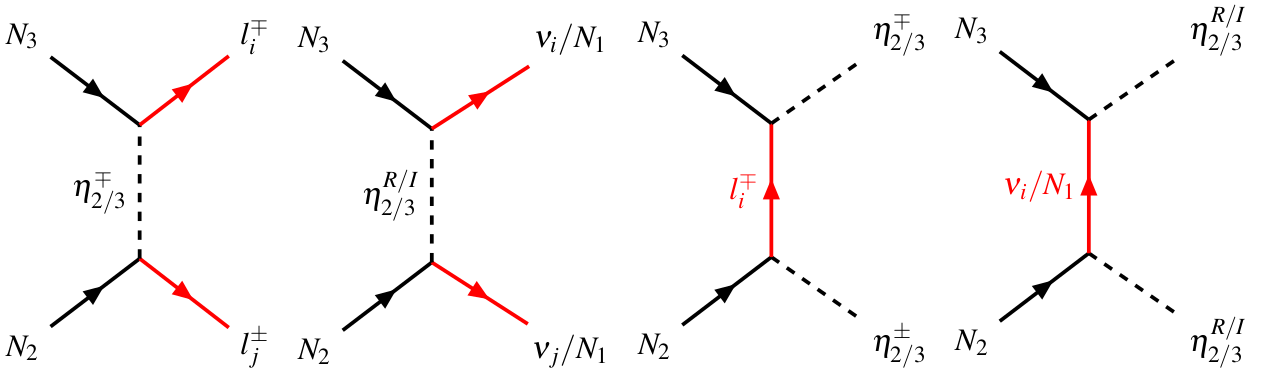}
\caption{\footnotesize \centering Fermionic DM co-annihilation channels involving both dark fermions.} 
\label{fig:coanni}
\end{figure}

\begin{figure}[h!]
\centering
\includegraphics[width=0.85\textwidth]{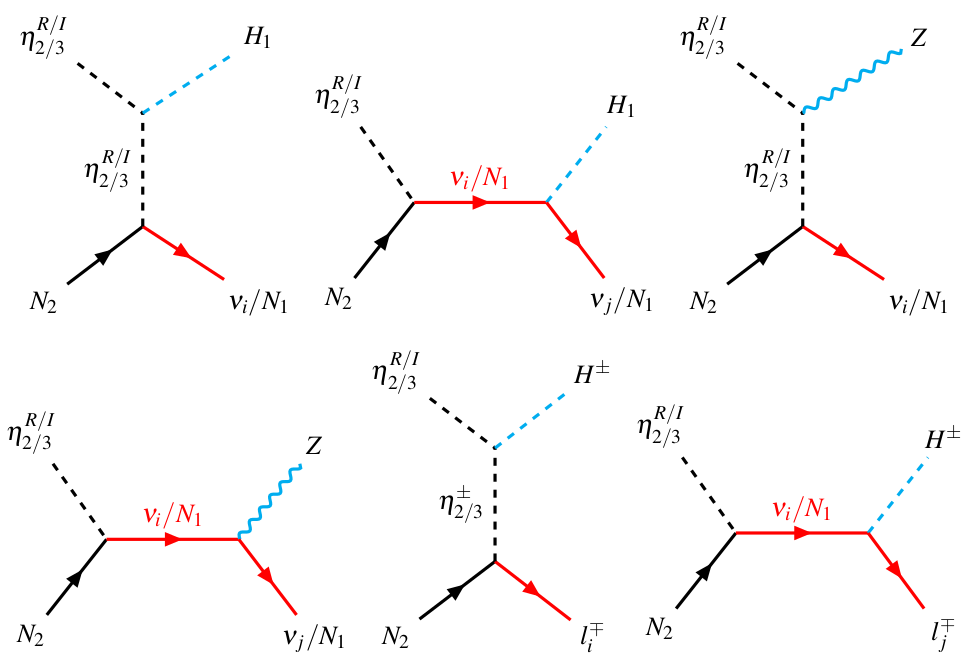}
\includegraphics[width=0.85\textwidth]{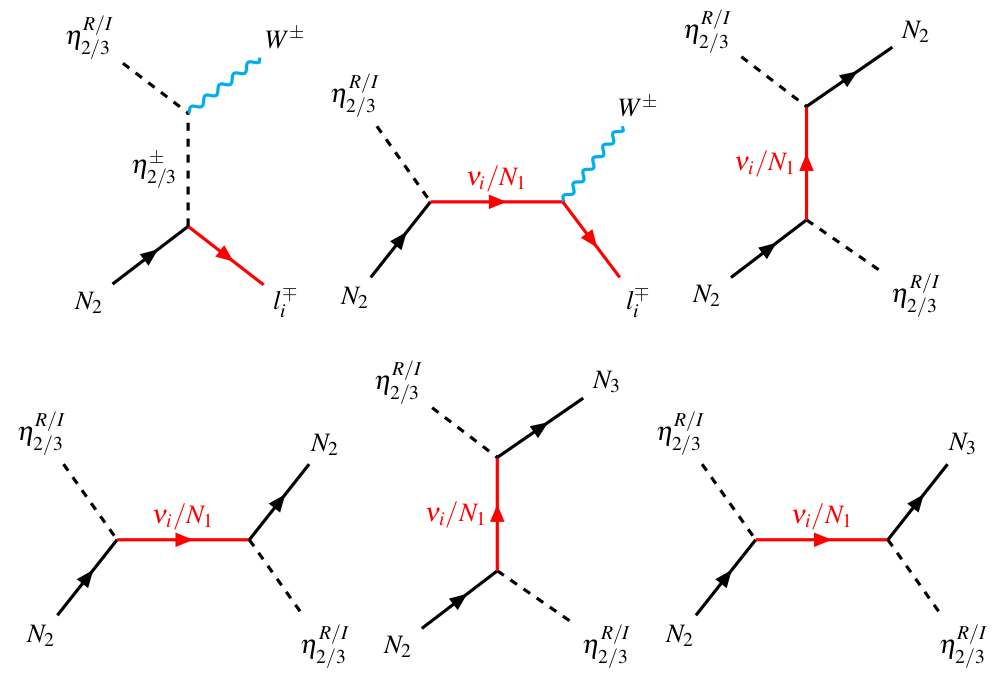}
\caption{\footnotesize \centering Fermionic DM co-annihilation channels involving neutral dark scalars.}  
\label{fig:coanni1}
\end{figure}

\begin{figure}[h!]
\centering
\includegraphics[width=0.85\textwidth]{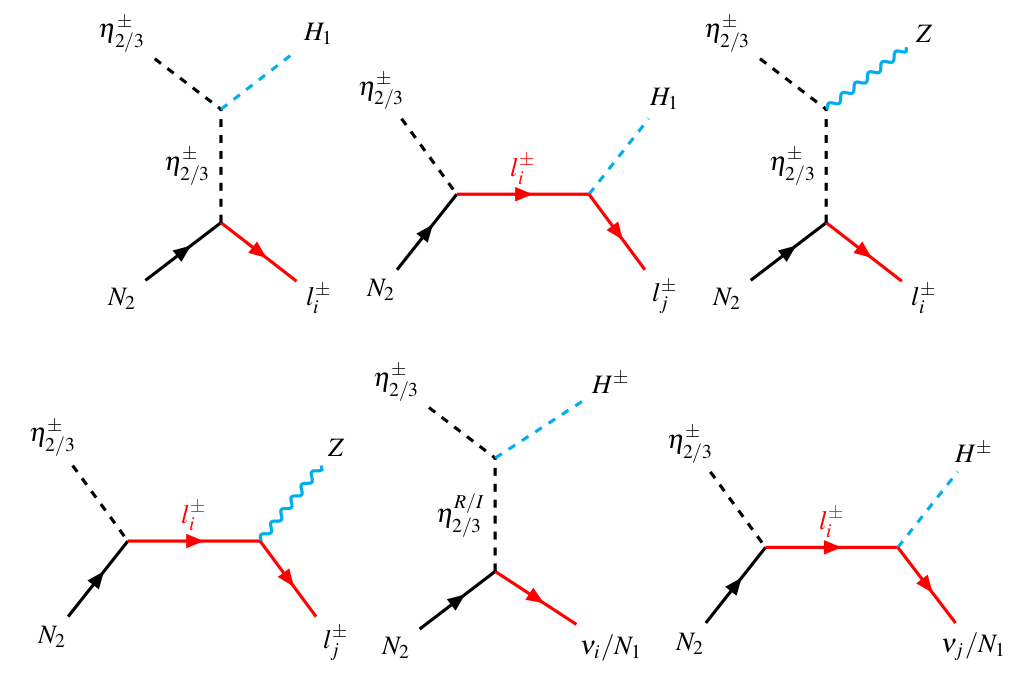}
\includegraphics[width=0.85\textwidth]{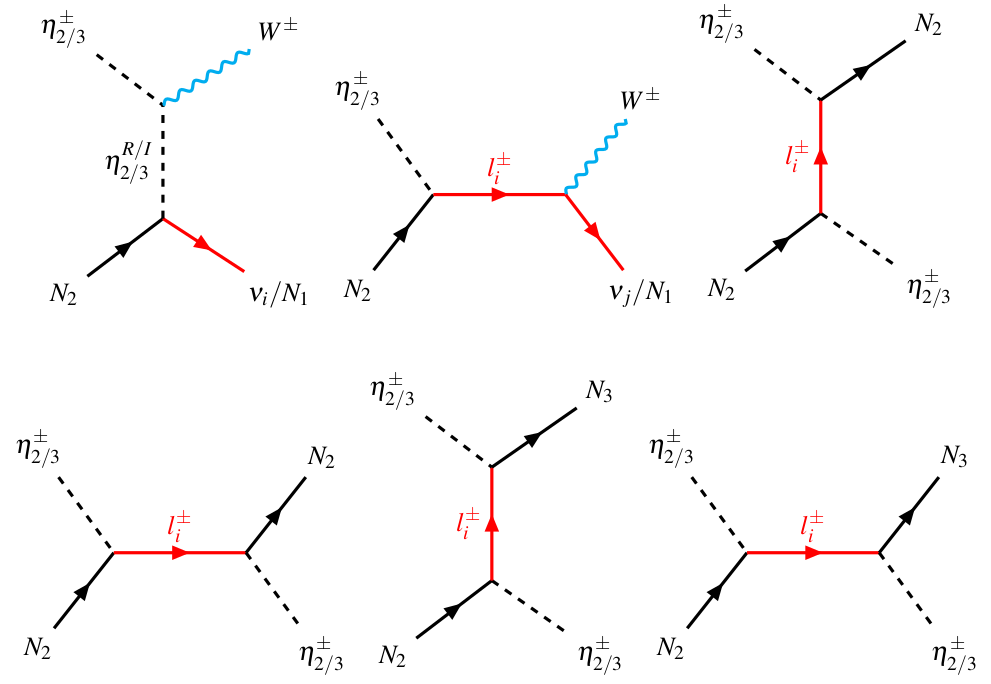}
\caption{\footnotesize \centering Fermionic DM co-annihilation channels involving charged dark scalars.}  
\label{fig:coanni2}
\end{figure}

\begin{figure}[t!]
\centering
\includegraphics[width=0.75\textwidth]{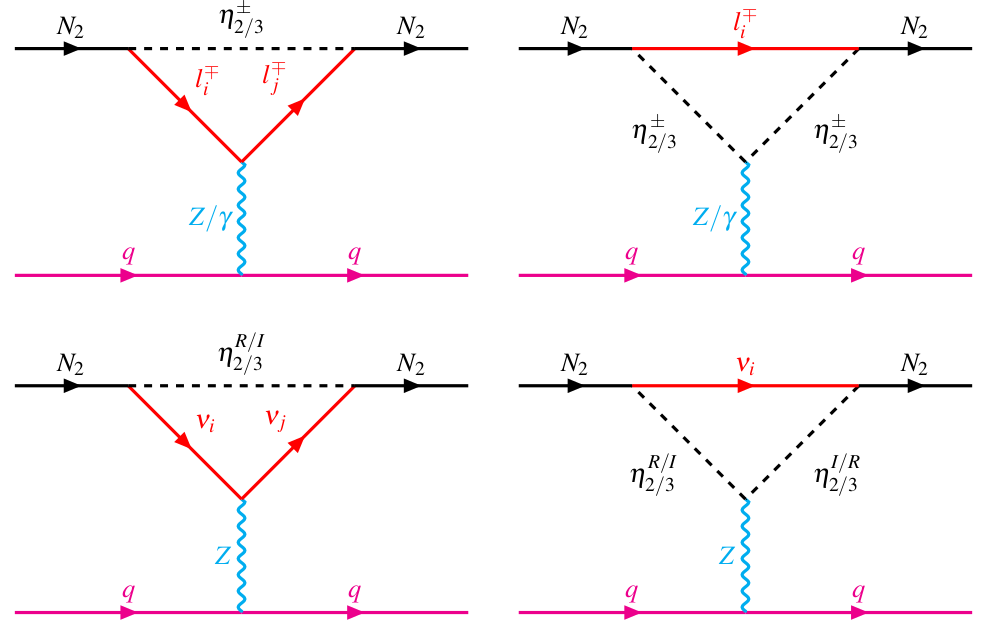}
\includegraphics[width=0.75\textwidth]{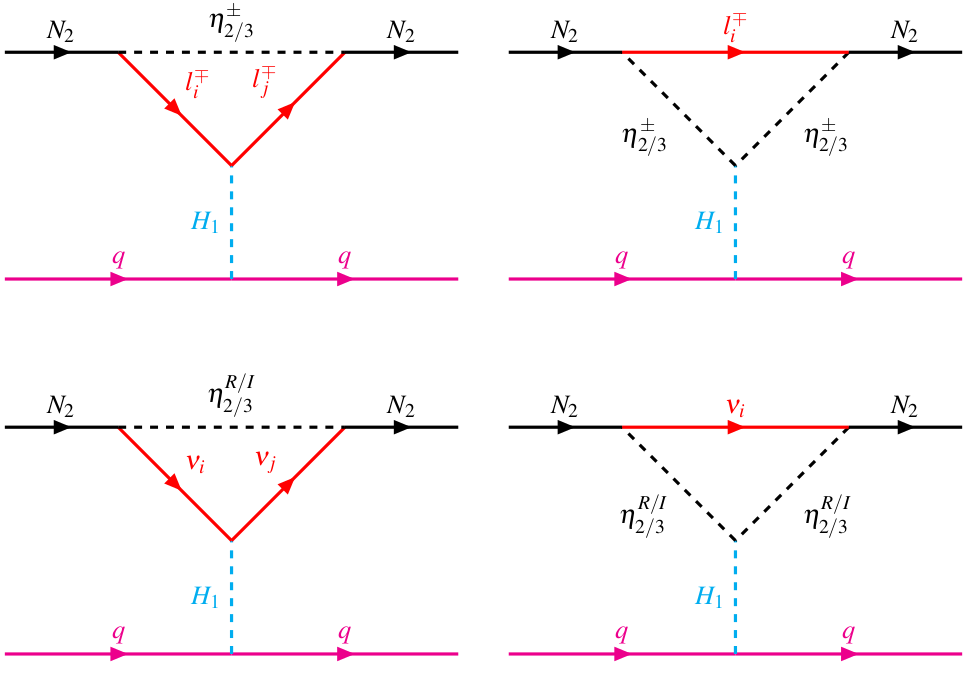}
\caption{\footnotesize \centering  Direct detection at one-loop level for fermionic DM.}  
\label{fig:DD}
\end{figure}

For the fermionic DM, the possible Feynman diagrams that govern the relic density are shown in Figs.~\ref{fig:anni1},~\ref{fig:coanni},~\ref{fig:coanni1} and ~\ref{fig:coanni2}. Furthermore, for the fermionic DM, we do not have tree-level interaction with quarks. Hence, Feynman diagrams for direct detection at the one-loop level are shown in Fig.~\ref{fig:DD}.
\vspace{0.10cm}
\FloatBarrier
\bibliographystyle{utphys}
\bibliography{bibliography.bib}             
\end{document}